\documentclass[11pt]{article}
\pdfoutput=1
\usepackage[margin=1.0in]{geometry}
\usepackage{amsmath,amssymb,graphicx}
\usepackage{hyperref}
\usepackage{slashed}
\usepackage{wasysym}

\usepackage[utf8]{inputenc}

\usepackage[greek,english]{babel}

 \usepackage[caption=false]{subfig}
\usepackage{booktabs}

\usepackage[font=small,labelfont=bf]{caption}
\usepackage{cite}
\usepackage{enumerate}

\newcommand{\iso}[2]{\mathcal{I}^\text{#1}_{#2}}

\newcommand{\efmin}{f_\text{min}}

\DeclareRobustCommand{\Sec}[1]{Sec.~\ref{#1}}

\DeclareRobustCommand{\App}[1]{App.~\ref{#1}}

\DeclareRobustCommand{\Fig}[1]{Fig.~\ref{#1}}

\DeclareRobustCommand{\Eq}[1]{Eq.~(\ref{#1})}
\DeclareRobustCommand{\Eqs}[2]{Eqs.~(\ref{#1}) and (\ref{#2})}
\DeclareRobustCommand{\Ref}[1]{Ref.~\cite{#1}}

\newcommand{\be}{\begin{equation}}
\newcommand{\ee}{\end{equation}}
\newcommand{\beq}{\begin{equation}}
\newcommand{\eeq}{\end{equation}}

\newcommand{\bea}{\begin{eqnarray}}
\newcommand{\eea}{\end{eqnarray}}
\newcommand{\half}{\frac{1}{2}}
\newcommand{\lmax}{\lambda_{\text{max}}}
\newcommand{\kref}{{\color{black}{\hat k}}}

\usepackage[table]{xcolor}
\usepackage{bm}

\definecolor{gray}{cmyk}{0,0,0,0.05}
\newcolumntype{a}{>{\columncolor{gray}} l}

\DeclareMathOperator{\tr}{tr}

\numberwithin{equation}{section}

\title{\bf The Efficacy of Event Isotropy as an Event Shape Observable}
\author{Cari Cesarotti,  Matthew Reece, and Matthew J.~Strassler \\
{\small Department of Physics, Harvard University, Cambridge, MA, 02138}}

\begin{document}
\maketitle

\begin{abstract}
Event isotropy $\iso{sph}{}$, an event shape observable that measures the distance of a final state from a spherically symmetric state, is designed for new physics signals that are far from QCD-like.  Using a new technique \cite{paper1} for producing a wide variety of signals that can range from near-spherical to jetty, we compare event isotropy to other observables.  We show that thrust $T$ and the $C$  parameter (and $\lmax$, the largest eigenvalue of the sphericity matrix) are strongly correlated and thus redundant, to a good approximation.     By contrast, event isotropy adds considerable information, often serving to break degeneracies between signals that would have almost identical $T$ and $C$ distributions.   Signals with broad distributions in $T$ (or $\lmax$) and in $\iso{sph}{}$ separately often have much narrower distributions, and are more easily distinguished, in the $({\iso{sph}{}},\lmax)$ plane.  An intuitive, semi-analytic estimation technique clarifies why this is the case and assists with the  interpretation of the distributions. 
\end{abstract}

%%%%%%%%%%%%%%%%%%%%%%%%%%%%%%%%%%%%%%%%%
\section{Introduction}
\label{sec:intro}
%%%%%%%%%%%%%%%%%%%%%%%%%%%%%%%%%%%%%%%%%

The Large Hadron Collider (LHC) and its detectors have delivered an enormous body of data, with much more to come in the future. So far, this has allowed a spectacular confirmation of Standard Model predictions, including not only the existence of the Higgs boson but the strength of several of its interactions. Despite this success, a daunting task remains, which will occupy particle physicists during the next two  decades: we must make effective use of the LHC's data to search for new physics. In recent years, this goal has spurred increasingly sophisticated efforts to not only search for well-understood models, but to mine data for subtle signals and to consider collider events from a variety of viewpoints.

One strategy for obtaining a fresh perspective on data is to develop new ways of characterizing events by defining new collider observables. In this paper, we will explore the use of a recently proposed new observable,  {\em event isotropy}, denoted $\iso{sph}{}$ \cite{Cesarotti:2020hwb}, albeit in the clean setting of an $e^+e^-$ collider. We will provide new evidence that this observable is a useful way to characterize a range of collider events that deviate strongly from typical QCD-like events (which contain a small number of collimated jets). The events that we will study interpolate between jetty events and quasi-spherical events (sometimes referred to as Soft Unclustered Energy Patterns or ``SUEPs'' \cite{Knapen:2016hky}).
%
%
%  for producing such a wide range of event shapes.

To explore the properties of these event shape observables, we generate a wide range of possible new-physics signals using a class of simplified models motivated by the Hidden Valley scenario, in which the physics of a hidden sector often leads to complex final states.\cite{Strassler:2006im, Strassler:2008bv} 
These simplified models, employing intuition from the AdS/CFT correspondence, can be viewed either as describing hidden sectors with an extra dimension (a slice of (4+1)d anti-de-Sitter space, as in Randall-Sundrum \cite{RandallSundrum:1999}) or, rather loosely, as describing the physics of a confining gauge theory which both resembles and yet differs from large-$N$ QCD.
The most relevant difference is that the gauge theory, rather than being asymptotically free, has a large `t Hooft coupling ($g^2 N\gg 1$ at all scales) while remaining approximately conformal in the ultraviolet.
In this regime, the extra-dimensional description, referred to as the ``hard wall'' model \cite{Polchinski:2001tt, Polchinski:2001ju, Polchinski:2002jw, erlich2005qcd, DaRold:2005mxj}, is straightforward.
This approach provides a simple, flexible framework for for creating unfamiliar collider physics signatures. 
 A detailed description of the hard wall model and our use of it is given in \Ref{paper1}.
We provide a brief summary here.

We use the hard wall model to describe a hidden sector with one or more towers of self-interacting particles.
These particles are Kaluza-Klein (KK) modes of the warped extra dimension, in which only SM-singlet fields propagate.
Motivated by the AdS/CFT correspondence for confining gauge theories, we will refer to these KK modes as ``hidden hadrons,'' or simply ``hadrons.''

Events of interest involve a collision of SM particles that produces a heavy hadron.
Implicitly, our studies focus on events at a hypothetical $e^+e^-$ collider that produces one such particle at rest, and nothing else.
This initial hadron decays into lighter hadrons, which decay in their turn, creating a cascade of hidden particles.
At the end of the cascade, we are left with many light hadrons, each of which is  stable against decays within the hidden sector; we refer to a hidden stable hadron as an HSH.

In a Hidden Valley model, some or all of these hadrons would decay to the SM, the details depending on the interaction between the visible and hidden sectors.
Since our focus is event shapes rather than on detailed phenomenology, we model this step by allowing each HSH to decay to two massless particles, which serve as proxies for SM particles.
We then treat these massless particles as our final state, and compute event-shape observables based upon them.

It has been known for some time that such Kaluza-Klein cascades can produce approximately spherical events \cite{csaki2009ads}, much as a CFT at large 't Hooft coupling is expected to do \cite{Strassler:2008bv, Polchinski:2002jw, Hatta:2008tx, Hatta:2008tn, Hofman:2008ar}.
 However, we have shown in \cite{paper1} (and in a preliminary form in Sec. 7.3 of \cite{Alimena:2019zri}), using novel analytic methods to compute the hadron self-couplings, that by changing bulk mass parameters or adding boundary-localized interactions, one can obtain a wide variety of  event shapes.  We exploit this variety in our study below.

The details of the calculations are as follows. 
 We  introduce an interacting scalar field theory in a five-dimensional ``bulk'' space, with coordinates $x^\mu,z$ and the AdS (anti-de Sitter)  metric.
If $z$ ran from 0 to $\infty$, the isometries of the AdS space would match those of a 4d conformal theory, but when we add a hard wall that forces $z<z_{\text{IR}}$, we break the isometry, and thus the  conformal invariance of any corresponding 4d theory.
With suitable boundary conditions, the spectrum of KK states resembles that of a confining gauge theory in 4d, with a mass gap and a tower of interacting hadrons.\footnote{Note this is only a toy model; a true confining and strongly-interacting gauge theory would necessarily have a  gravitating bulk with an infinite tower of 5d fields, the equivalent of ten bulk dimensions.}

For simplicity we consider only scalar fields in the bulk.
Each field $\Phi$ of 5d mass $M$ produces an infinite Kaluza-Klein tower of states with wavefunctions that factorize into 4d scalar modes $\phi_n$ and wavefunctions $\psi_n(z)$ of the compact fifth dimension 
\begin{equation}
\Phi\left(x^\mu, z \right) = \sum_{n=1}^\infty \phi_n \left( x^\mu \right) \psi_n \left(z\right) 
\end{equation}
where
\begin{equation}
\psi_n(z) \propto z^2 J_\nu(m_n z) \ .
\end{equation}
We often refer to the mode number $n$ as the \textit{KK-number}.
These modes are loosely interpreted, within a corresponding 4d theory, as a tower of hidden scalar hadrons sourced by a field-theory operator ${\cal O}$.  This operator must have scaling dimension $d_{\cal O} \equiv \nu + 2$,
where
\begin{equation}
\nu \equiv \sqrt{4+M^2R^2},
 \label{eq:nuFromMass}
\end{equation}
and $R$ is the AdS curvature radius.

Cubic interactions $\Phi_1 \Phi_2 \Phi_3$ in the 5d scalar theory result in infinitely many interaction terms among the 4d hadrons, where the couplings are set by the overlap of the wavefunctions along the warped extra dimension.
These interactions allow heavier hadrons to decay to lighter ones, inducing the desired cascades.
%a mode high in the tower (i.e., with large KK-number), when produced at a collider, to decay into other hadrons.  These in turn decay into more hadrons, and a cascade ensues.  At the end of the hadronic cascade, the state consists of a large number of light hadrons.  These SM-singlet particles subsequently decay back to observable Standard Model particles.  What results is a hidden-valley signature, whose details depend on the interactions between the visible and hidden sectors, but whose kinematic features, on which we focus here, are less model-dependent, as long as all or most of the energy enters the detectors in a visible form.
%

We consider three scenarios, described in detail in \cite{paper1}:
\begin{itemize}
\item The ``single field'' case, which involves a cubic self-coupling $\Phi_1^3$;
\item The ``two-field'' case, where two fields couple via an interaction\footnote{We do not include other terms; those odd in $\Phi_2$ can be forbidden with a symmetry, and there are physically motivated situations where $\Phi_1^3$ terms are small or absent.} $\Phi_1 \Phi_2^2$;
\item The ``bulk-boundary'' case, where a single field has a coupling at the IR boundary of the warped dimension $\Phi_1^3\delta(z-z_\text{IR}) $ which may compete with a bulk coupling $\Phi_1^3$.
\end{itemize}
 In the first two scenarios, we consider Dirichlet boundary conditions: $\Phi_i(z_{\rm IR}) = 0$. In the case with boundary coupling, we consider Neumann boundary conditions: $\partial_z \Phi_1(z_{\rm IR}) = 0$. 
 For the single field case, we study $\nu=0,0.15,0.75$; in the two-field case  we take $\nu_1=2,3,4$ and $\nu_2=0$; and for the bulk-boundary case, we use $\nu=0.3$.

In the single-field scenario, we found in \cite{paper1} that for physically motivated values of $\nu$, the pattern of hadronic couplings favors decays where the sum of the KK-numbers of the daughters is as close to the parent's KK-number as is kinematically allowed. 
Thus KK-number is approximately conserved.
Conservation is strongest at the Breitenlohner-Freedman bound of $\nu = 0$ \cite{Breitenlohner:1982jf} and is increasingly violated for larger values of $\nu$.
%
%We refer to this approximate symmetry as \textit{KK-number conservation}.
%
Since the mass of a mode is nearly linear in the KK-number, approximate KK-number conservation implies that most decays are near kinematic threshold,  in which case the cascade develops without producing many boosted particles, and the final state of massless particles is quasi-isotropic.

In the two-field scenario, the event shape is characterized primarily by the difference $\nu_1 -2 \nu_2$.
For values $\nu_1 > 2\nu_2+2$, KK-number is no longer approximately conserved.
In the space of couplings, there are regions of unsuppressed KK-number violation, which we referred to as `plateaus' in \cite{paper1}.
The number of plateaus increases as $\nu_1$ increases for a fixed, small $\nu_2$. 
In a decay with large KK-number violation, the daughters are far from kinematic threshold and can be produced with a substantial boost.
Subsequent decays of a highly boosted particle form a jet.
Thus  KK-number violation leads to events with a highly anisotropic, often jetty radiation pattern. 

Finally, for a boundary coupling, KK-number violation can be even stronger.
If the boundary coupling dominates over any bulk interaction, then 
couplings among all triplets of hadrons are approximately equal, because
the magnitude of their wavefunctions are approximately equal [$\psi_i (z_\text{IR}) \approx \psi_j (z_\text{IR})$] at the boundary.
The cascades in this scenario generate  jetty, highly anisotropic events.

%\ms{This paragraph is really appropriate to an older version; needs more work still.} \cjc{Not editing this paragraph} 
In this paper we use simulated events in the scenarios above to study $\iso{sph}{}$ in more detail, and to argue that $\iso{sph}{}$ provides a useful measure of event properties which is  distinct from information captured by certain traditional observables, such as thrust and the $C$-parameter, that are effective measures of the proximity of an event to a dijet configuration.
%any information captured by \ms{This claim is too strong; we do not prove something so general, only something about thrust and multiplicities} more traditional event observables. 
To this end, in \Sec{sec:EventShape} we review a collection of well-established ways to characterize events, including thrust and eigenvalues of the generalized sphericity tensor. In \Sec{sec:JM}, we examine how event isotropy is complementary to measures of particle or jet multiplicities.
%, and discuss why we might wish to fix the multiplicity before evaluating the value of event isotropy. %motivate a particular choice to count the number of events above a fixed fraction of the total event energy as a useful observable for our purposes. 
In \Sec{sec:SimpleModel}, in order to gain intuition for how event isotropy works and how it differs from other known variables, we estimate event shape observables in some simple final states. These estimates allow us to map out how collider events might populate the two-dimensional $(\iso{sph}{}, \lmax)$ plane, where $\lmax$ is the largest eigenvalue of the sphericity tensor [see  \Eqs{eq:spherTensor}{eq:lambdaMax}]. Finally we turn in \Sec{sec:simResults} to the study of events generated by the 5d simplified models of \cite{paper1}. We conclude in \Sec{sec:discussion} and point out some open questions for future studies. %with a discussion on how event isotropy might be used along with other event shape observables for future studies and searches at colliders. \ms{do we really do this?}

%We begin with simple models with a single field with a bulk interaction. These models generate events that are approximately, but not perfectly, spherical. We provide an approximate analytic understanding of how deviations from idealized isotropy affect the numerical value of $\iso{sph}{}$. 
%%%%%%%%%%%%%%%%%%%%%%%%%%%%%%%%%%%%%%%%%
\section{Event Shape Observables}
\label{sec:EventShape}
%%%%%%%%%%%%%%%%%%%%%%%%%%%%%%%%%%%%%%%%%
\subsection{Definition of Observables}

In this section, we will review the event shape observables that are relevant for our studies in subsequent sections. We consider the sphericity tensor \cite{Donoghue:1979vi,Ellis:1980wv}, thrust \cite{Brandt:1964sa,Farhi:1977sg,DeRujula:1978vmq}, and the  recently proposed event isotropy $\iso{sph}{}$ \cite{Cesarotti:2020hwb}. These observables are sensitive to different features in the event shape. We will focus our attention on observables designed for $e^+ e^-$ events, i.e., observables with spherical rather than cylindrical symmetry, defined in the center-of-momentum frame. This choice will allow us to understand the new event isotropy variable $\iso{sph}{}$ in an idealized setting, before introducing all of the complications associated with hadron colliders. 

The generalized sphericity tensor is defined as
\begin{equation}
S^{(r)ij} = \frac{\sum_m |p_m|^{r-2} p_m^i p_m^j}{\sum_m |p_m|^r},
\label{eq:spherTensor}
\end{equation}
where $m$ runs over the number of particles in the event and $i, j$ are 3-vector indices. We will limit our attention to the case $r = 1$, for which the eigenvalues of this tensor are infrared and collinear safe (``IRC-safe'') observables.\footnote{The sphericity tensor in the  case $r = 2$ is used to define the observable simply known as sphericity \cite{Bjorken:1969wi,Parisi:1978eg}, which has been used in earlier related studies \cite{csaki2009ads,Alimena:2019zri}. However, sphericity is not an IRC-safe observable, and we do not consider it further in this paper.} 
For an event with $k$ particles, the sphericity tensor can be computed in $\mathcal{O}(k)$ time.

By construction, $\tr S^{(r)} = 1$, so only two eigenvalues are independent.
% Hence the three eigenvalues sum to 1, and any two independent combinations of the three eigenvalues contain all of the invariant physical information captured by the tensor. 
In the literature, two independent combinations of eigenvalues known as the $C$-parameter and the $D$-parameter have commonly been used:
\begin{align}
C &= 3 \left( \lambda_1 \lambda_2 + \lambda_1 \lambda_3 + \lambda_2 \lambda_3 \right), \label{eq:cParmDef} \\ 
D &= 27 \lambda_1 \lambda_2 \lambda_3 \ ;
\label{eq:dParmDef} 
\end{align}
the normalizations assure both variables run between 0 and 1.
For dijet events, both $C$ and $D$ are zero. For three-jet events, which are planar, one eigenvalue is zero, so $D = 0$ but $C \neq 0$. Hence $D$ measures a deviation from coplanarity, whereas $C$ measures a deviation from collinearity \cite{Donoghue:1979vi, Ellis:1980wv}.

%In this case, we find that the largest eigenvalue of $S^{(1)}$, which we will denote $\lmax$, is a useful observable. 
 However, the events that we study in this paper are far from being either collinear or coplanar. 
For our purposes, the maximum eigenvalue of the sphericity tensor
\begin{equation}
\lambda_\text{max} = \max\{\lambda_1, \lambda_2,\lambda_3\}.
\label{eq:lambdaMax}
\end{equation}
is more useful.
 Within the event samples that we study, the values of $C$, $D$, and $\lmax$ are highly correlated, but $\lmax$ has simpler analytic properties. It lies in the range
\begin{equation}
\frac{1}{3} \leq \lmax \leq 1.
\end{equation}
The upper bound is achieved when the two smaller eigenvalues are zero, i.e., for dijet events. Spherical events have $\lmax = 1/3$.   But $\lmax$ is not a true measure of isotropy, because this lower bound is saturated for any triaxially symmetric distribution (symmetric under all exchanges of axes), such as the configuration of six particles with equal and orthogonal momenta oriented along the positive and negative axes. 

Thrust, by contrast, is a true measure of isotropy, as it is maximized for two particle back-to-back events and minimized only for spherically uniform radiation patterns.\footnote{A proof of this claim is given in \Ref{Cesarotti:2020hwb}.}
Defined as 
\begin{equation}
\label{eq:thrust_def}
T(\mathcal{E}) = \max_{\hat{n}} \frac{\sum_i | \vec{p}_i \cdot \hat{n}| }{\sum_j |\vec{p}_j|},
\end{equation} 
where $\hat{n}$ is chosen to maximize $T$ and is known as the thrust axis,
thrust ranges from $T=0.5$ for an isotropic event to $T=1$ for two back-to-back particles.
A downside of this variable is the computationally expensive minimization procedure for events with high multiplicity $k$. 
Not only does the computation time scale as  $\mathcal{O}\left(k^2 \log k \right)$  (a recent improvement \cite{Wei:2019rqy} upon earlier algorithms which have $\mathcal{O}\left(k^3\right)$ scaling), 
the overall coefficient is large.  
We find that even when using the improved algorithm, the computation time for the thrust of an event with $k\sim100$ is several seconds. 

Another shortcoming of thrust for complex events is its small dynamic range in the near-spherical regime. 
Thrust is most sensitive to deviations from the dijet configuration and is less effective at discriminating isotropic samples from multijet events.\footnote{An observable that was introduced for $pp$ colliders specifically to enhance the sensitivity in deviations of complex events was \textit{supersphero} \cite{Banfi:2010xy}. 
However, there are as yet no tools for computing any generalization of this variable to  $e^+e^-$ colliders. }

Meanwhile, event isotropy is a measure of the distance from a uniform radiation pattern.
It is a novel application of the concept of Energy Mover's Distance (EMD) \cite{Komiske:2019fks,Komiske:2020qhg}. 
The EMD is the minimum ``work''  necessary to rearrange one radiation pattern of massless particles $P$ into another radiation pattern $Q$. 
The radiation pattern is defined as a collection of particles, each of which is specified by its position on the unit sphere and fraction of the total energy. 
A given rearrangement of $P$ into $Q$ is specified by the transport map $f_{ij}$ which tracks the fraction of the total event energy moved from position $i$ to position $j$.
A measure of the work done by this rearrangement is the sum of the distance $d_{ij}$ from $i$ to $j$, weighted by the fraction of energy $f_{ij}$.
The EMD is defined to be the minimum work, for all possible rearrangements from $P$ to $Q$:
\begin{equation}
\text{EMD}\left( P, Q \right) = \min_{\{f_{ij}\}} \sum_{ij} f_{ij} d_{ij}. 
\label{eq:EMDdef}
\end{equation}

To compute the EMD, one must specify the distance measure $d_{ij}$. Here, as in \cite{paper1} (but in contrast to \Ref{Cesarotti:2020hwb}), we take  the distance between two massless particles $i,j$ to be
%in the distributions is
%
\begin{align}
d_{ij} %&= \frac{3}{2}\sqrt{ \hat{n}^\mu_i  \hat{n}_{\mu,j} } \nonumber \\
& = \frac{3}{2}\sqrt{ 1 - \cos \theta_{ij} },
\label{eq:distMeas}
\end{align}
where $\theta_{ij}$ is the angle between their three-momenta.
This is a true metric, satisfying the triangle inequality. 
The normalization is chosen such that the two-particle back-to-back configuration has $\mathcal{I}^\text{sph} = 1$ and a perfectly uniform spherical configuration has $\mathcal{I}^\text{sph} = 0$. 
Note that since the event isotropy measures the distance between radiation patterns, a {\it smaller} value of $\iso{sph}{}$ corresponds to a {\it more} isotropic configuration. 

Because the EMD is linear in energy fractions, it has a useful convexity property which event isotropy inherits. If we consider two radiation patterns $P$ and $P'$ with the same total energy, the EMD of a weighted sum $x P + (1-x) P'$ obeys 
\begin{equation}
\text{EMD}(xP + (1-x) P',Q) \leq x\, \text{EMD}(P,Q) + (1-x)\, \text{EMD}(P',Q),
  \label{eq:EMDconvexity}
\end{equation}
 because it is always possible to move the combined energy by rearranging the energy of $P$ and $P'$ independently.

The event isotropy of an event $\mathcal{E}$ appropriate to an $e^+e^-$ collider is the EMD from the event to a reference sphere.
Ideally the reference sphere would be perfectly smooth, but in practice it must be discretized. 
Define $\mathcal{U}^\text{sph}_\kref$ to be a set of $\kref$ points symmetrically distributed on a sphere.  
Then the event isotropy with respect to a $\kref$-pixel sphere is
\begin{equation}
\iso{sph}{\kref} \left(\mathcal{E}\right)= \text{EMD}\left( \mathcal{U}^\text{sph}_\kref, \mathcal{E}\right).
\label{eq:EIdef}
\end{equation}
Convenient finite-multiplicity pixelated reference spheres can be generated using the \texttt{HEALPix} Python framework \cite{2005ApJ...622..759G}.
Unless otherwise specified,  we will take $\kref$=192. 

The fact that $\kref$ is finite has important effects that would be absent in the $\kref\to\infty$ limit.  For instance, if we take a set of  events that consist of  $k$ particles of equal energy distributed uniformly on the sphere, but with each event oriented randomly, and compute their average event isotropy with respect to a reference sphere of $\kref$ pixels, then 
\begin{equation}\label{eq:theoryminimum}
\iso{sph}{\kref}(\mathcal{U}_{k}^\text{sph}) \approx \sqrt{2/{\text{min}}(k,\kref)}
\end{equation}
as shown in \App{ap:EMD}, \Eqs{eq:approx0}{eq:approx}.
Since less isotropic events have larger average event isotropy than spherical events, this sets a theoretical miminum for $\iso{sph}{\kref}$ in samples of events with $k$ particles.  The fact that $\iso{sph}{192}$ can differ from $\iso{sph}{\infty}$ by as much as $\sqrt{2/192}\sim 0.1$ in some cases has a minor but noticeable impact on our results. 
Further discussion of this and related issues is presented in \App{ap:EMD}.

The worst-case computation time of event isotropy scales as $\mathcal{O}(k^3 \log^2[k])$ for events with $k$ particles. 
Although this is parametrically worse than thrust, the overall coefficient is much smaller: a $\sim$100-particle event with a 192-pixel reference sphere takes approximately 50 milliseconds to compute.

%
%%%%%%%%%%%%%%%%%%%%%%%%%%%%%%%%%%%%%%%%%
\subsection{Correlations among Observables}

In this section we remark on correlations between thrust $T$ and the observables built from the sphericity tensor $S$ known as the $C$ and $D$ parameters, and illustrate them using the events from our simulations.
It is easy to see that these observables are tightly correlated in events with azimuthal  symmetry around a central axis.
Violation of this symmetry breaks the correlations, but does so rather slowly, especially for $T$ and $C$, and in practice the correlations in the symmetric limit are present in a large fraction of the events in many signals.

Let us estimate the degree of correlation, in the azimuthally symmetric limit, between thrust and $\lmax$, defined in \Eq{eq:lambdaMax}.
Without loss of generality we take the axis of symmetry to be the $z$ axis and define $\theta$ as the polar angle. 
When a majority of the momentum is aligned along the axis of symmetry, the thrust axis and the eigenvector corresponding to $\lmax$ lie along the $z$ axis, in which case 
\begin{equation}
T = \frac{\sum_i |\vec{p}_i \cdot \hat{z}|}{\sum_i |\vec{p}_i|} \ ;
\end{equation}
\begin{equation}
\lambda_\text{max} = \frac{\sum_i (\vec{p}\cdot \hat{z})^2/|\vec{p}_i|}{\sum_i |\vec{p}_i|} \ .
\end{equation}
 The other eigenvalues of the sphericity matrix are both $\frac{1}{2} (1-\lambda_\text{max})$ due to the property $\tr S^{(1)}=1$ and to azimuthal symmetry, and so for these events $C$ and $D$ are determined by $\lmax$.
%Thus we need not discuss the $C$- and $D$-parameter for these event configurations as they are exactly correlated with $\lambda_\text{max}$. 

\begin{figure}[t!]
\centering
\subfloat[]{
      \includegraphics[width=0.44\textwidth]{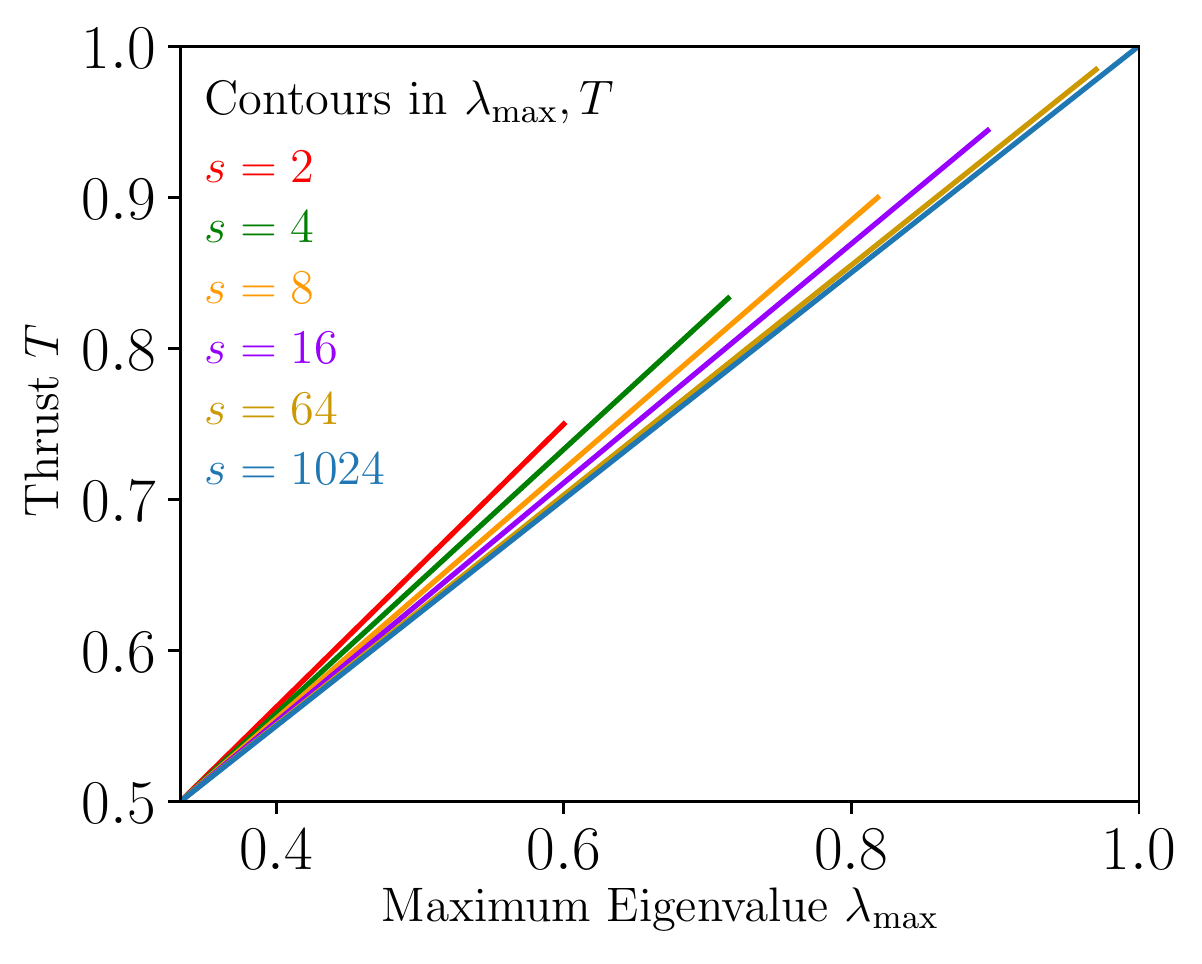}
     }
\hfill
\subfloat[]{
      \includegraphics[width=0.46\textwidth]{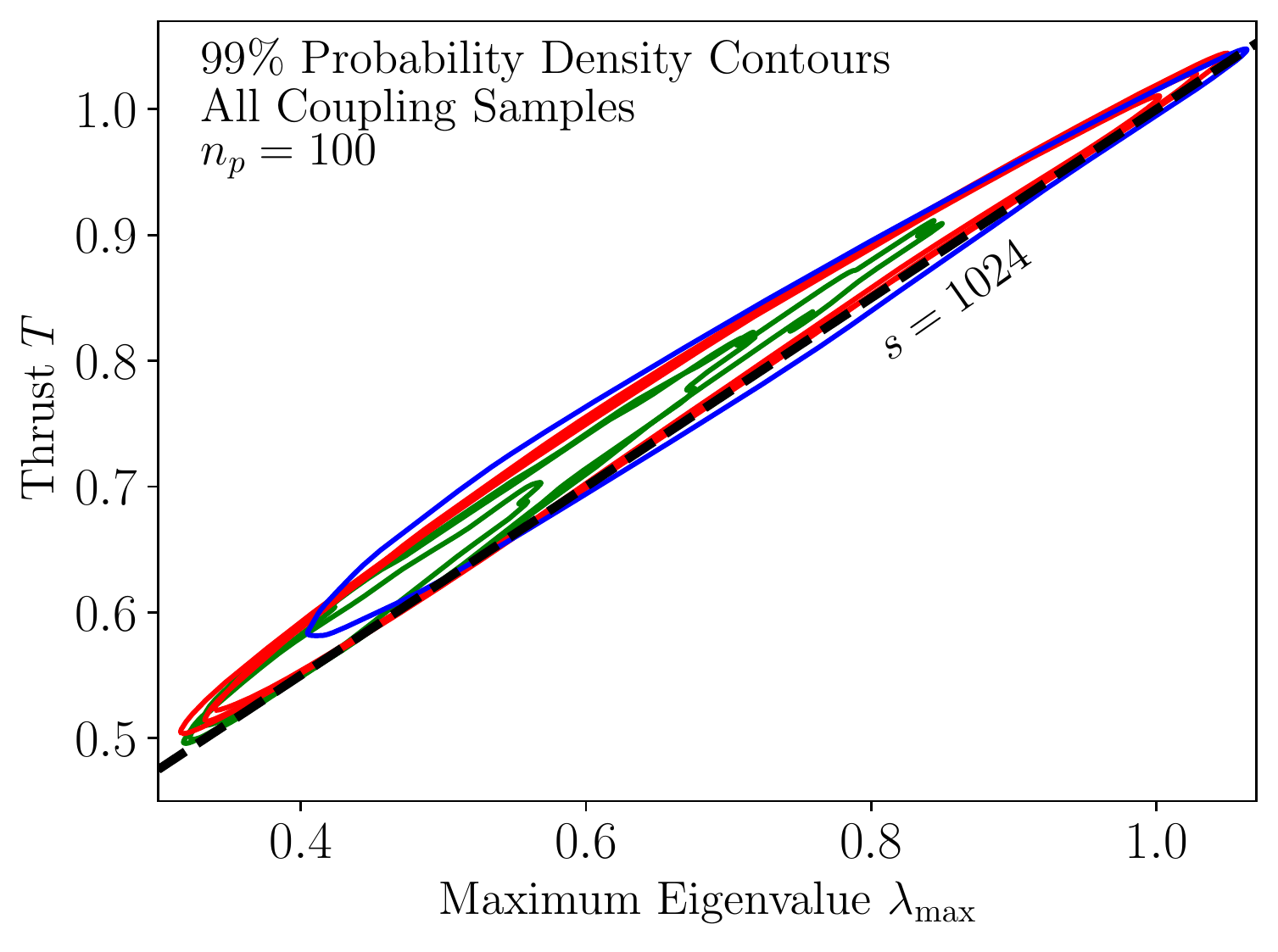}
     }
\caption{(a) For the azimuthally-symmetric radiation patterns in \Eq{eq:probDist}, the relation between $T$, \Eq{eq:P_T}, and $\lambda_\text{max}$, \Eq{eq:P_lmax}, is shown for various fixed values of $s$ and a large range of $\epsilon$.  The correlation between thrust and the maximum eigenvalue is persistent, as the contours have similar characteristic slopes. (b) Contours enclosing 99\% of the events in our simulated samples are shown for the single field samples (green), the two-field samples (red), and the boundary sample (blue). To guide the eye, the $s=1024$ curve from (a) is shown as a dashed line.  Although our simulated events are not azimuthally symmetric, the region they populate is well-modeled by the curves in (a). }
\label{fig:tvslambdamax}
\end{figure}

Let us consider classes of events with azimuthal symmetry that consist of a radiation pattern with energy density 
\begin{equation}
\label{eq:probDist}
P(\theta) \propto  ( 1 + \epsilon \cos^s \theta ) , 
\end{equation}
where $\epsilon>0$ and $s$ is an even integer. 
With the parameters $s$, $\epsilon$, we can interpolate between the spherical limit ($\epsilon \rightarrow 0$, or $s\rightarrow 0$, or $\epsilon$ finite with $s\rightarrow \infty$) and the dijet limit ($\epsilon \rightarrow \infty$ followed by $s \rightarrow \infty$). 
For these distributions, 
\begin{equation}
\label{eq:P_T}
T = \frac{\int {\rm d} \Omega \left| \cos \theta\right| \left(1 + \epsilon \cos^s \theta \right)}{\int {\rm d} \Omega \left(1 + \epsilon \cos^s \theta \right) } = \frac{\frac{1}{2} + \frac{\epsilon}{2+s}}{1 + \frac{\epsilon}{1+s}} ;
\end{equation}
\begin{equation}
\label{eq:P_lmax}
\lambda_\text{max} = \frac{\int {\rm d} \Omega \cos^2 \theta \left(1 + \epsilon \cos^s \theta \right)}{\int {\rm d} \Omega \left(1 + \epsilon \cos^s \theta \right) } = \frac{\frac{1}{3} + \frac{\epsilon}{3+s}}{1 + \frac{\epsilon}{1+s}}.
\end{equation}
Their ratio,
\begin{equation}
\frac{T}{\lambda_\text{max}} = \frac{\frac{1}{2} + \frac{\epsilon}{2+s}}{\frac{1}{3} + \frac{\epsilon}{3+s}} \ ,
\label{eq:TlambdaRatio}
\end{equation}
is nearly constant over the allowed range  of $\epsilon$ and $s$.
To illustrate this, we plot $T$ and $\lmax$ in \Fig{fig:tvslambdamax}a, for several values of $s$, while varying $\epsilon$ from zero to infinity. 
The lines are of finite length as $P(\theta)$ becomes independent of $\epsilon$ as $\epsilon\to\infty$.
Because the lines have a common origin and similar slopes, their envelope forms a narrow band.\footnote{This remains true for all angularities \cite{Berger:2003iw} with negative semi-definite parameter.  As the parameter is taken positive, the band rapidly widens and the correlation is lost.  We have not explored the potential use of angularities any further.}
%%%%%%%%%%%%%%%%%%%%%%%%%%%%%%%%%%%%%
%%%%%%%%%%%%%%%%%%%%%%%%%%%%%%%%%%%%%%
%

Although this band is derived for signals whose events have azimuthal symmetry, it holds to a good approximation even when azimuthal symmetry is routinely violated by moderate (but not maximal) amounts, as commonly occurs in complex events. 
We can see this quantitatively by examining the event shapes of the simulations produced with our toy model, whose events are not generally azimuthally symmetric. 
For each of our simulations, grouped by category (single-field, two-field, or boundary; see \Sec{sec:intro}), the contours containing 99\% of the events from that simulation are plotted in \Fig{fig:tvslambdamax}b.
The envelope of these contours neatly outlines the same region predicted in \Fig{fig:tvslambdamax}a. 

Because of this very strong correlation, {\it we will use the maximum eigenvalue $\lmax$ in lieu of thrust for the remainder of this paper}.
The eigenvalue is much faster to compute, and is more tractable analytically. 

In a similar way, the fact that $\lmax$ determines $C$ and $D$ in the azimuthally-symmetric limit is still approximately true even when the symmetry is somewhat violated.
This is especially true for the $C$ parameter.
%Now let us turn to the $C$ and $D$ parameters, which are determined by the two independent eigenvalues of the $3\times 3$ sphericity tensor, which has trace 1.  
%
If we take $\lambda_1,\lambda_2$ as the two smallest eigenvalues of the sphericity tensor $S$, we may define $\lambda_-=\frac12 |\lambda_1-\lambda_2|$, which satisfies
\begin{equation}
0 \leq \lambda_- \leq {\rm min}\left[\frac12 \bigg(1-\lmax\bigg),\frac32 \left(\lmax - \frac13\right)\right] \ .
\end{equation}
If azimuthal symmetry (or merely symmetry involving rotations by 90 degrees around the $z$ axis) is preserved, $\lambda_-=0$.  Since only two of the eigenvalues of $S$ are independent, we may express $C$ and $D$ in terms of $\lmax$ and $\lambda_-$. 
 For instance, 
\begin{equation}
C = 3 \left(\frac{1}{4} (1 + 3 \lmax)(1 - \lmax) - \lambda_-^2\right) < \hat C \equiv C|_{\lambda_-=0}
%3\left(\lmax(1-\lmax)+\frac{1}{4}\lmax(1-\lmax)^2\right
\ .
\end{equation}
Inspection shows that $C$ never can differ much from its maximum value $\hat C$, and does so only in the vicinity of $\lmax\sim \frac12$.  Thus to a good approximation $C\approx \hat C$ is a nonlinear function of $\lmax$, and may be replaced with the latter.

\begin{figure}[!h]
\centering
\includegraphics[width=0.45\textwidth]{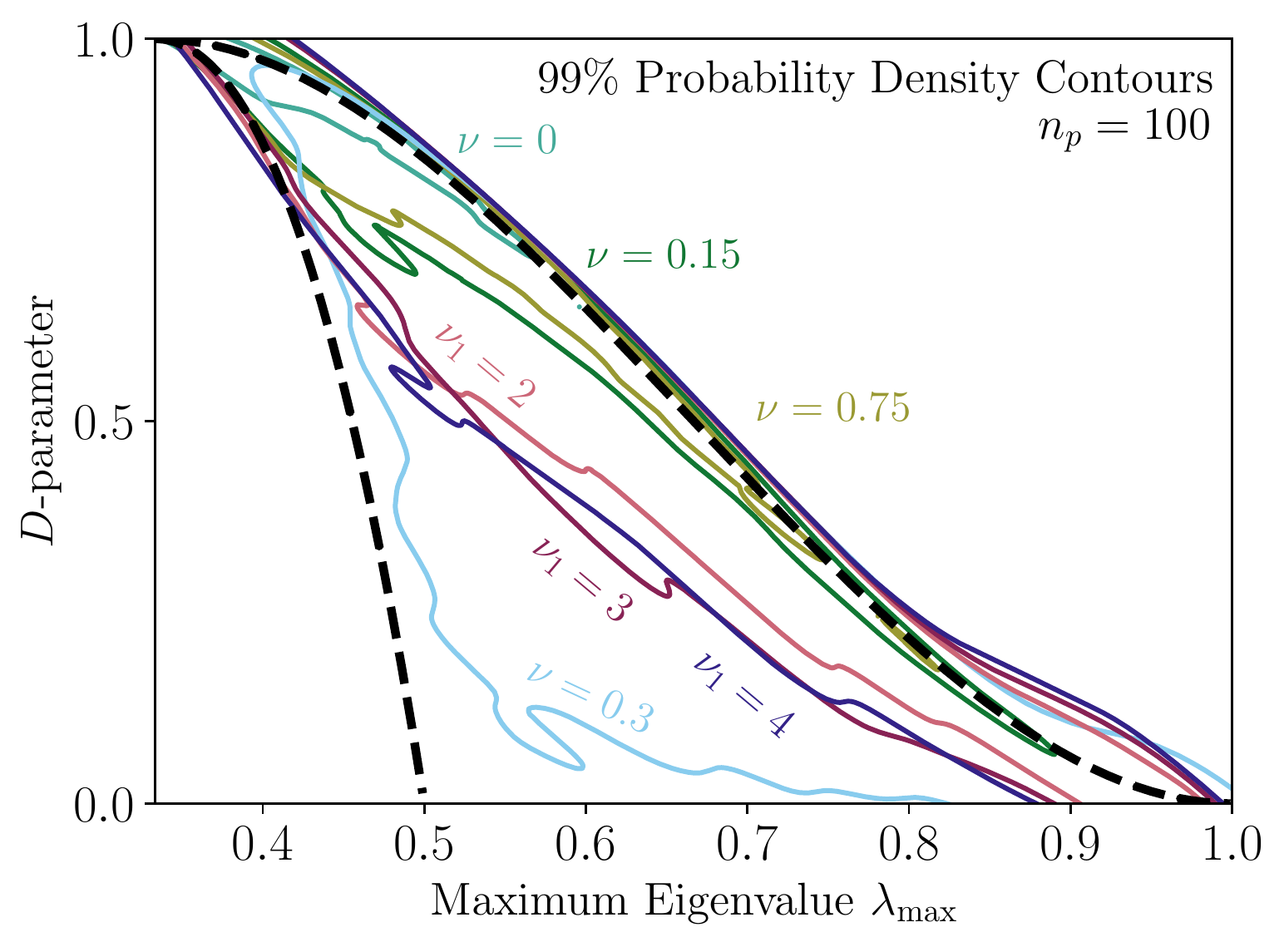}
\caption{Contours enclosing 99\% of the events in our simulated samples are shown in the plane of $D$ vs.~$\lmax$.  The mathematically-allowed region is bounded by the dashed curve.  (A plotting artifact causes contours to cross this curve, though no events actually do.) The full allowed region is accessed, though the lower left region is covered only by the tail of one distribution. Contours of 75\% or 50\% (not shown) are more restricted, lying near the upper edge of the allowed region. }
\label{fig:DvsLmax}
\end{figure}

The $D$ parameter, which in the above variables is
\begin{equation}
 D = 27\lmax \left(\frac{1}{4}(1-\lmax)^2 -\lambda_-^2\right) <  \hat D \equiv D|_{\lambda_-=0}
 %27\lmax \left(\frac{1}{4}\lmax(1-\lmax)^2\right)
 \  ,
\end{equation}
is considerably more independent.  In signals with relatively few particles or jets, including QCD itself, $D$ can lie far below $\hat D$; in planar events, it vanishes.  In more complex signals with many jets, however, events are typically far from planar, $\lambda_-$ tends to be small, and so $D$ tends to be closer to $\hat D$, in which case it is relatively correlated with $\lmax$.    The distributions of our simulated events in the $(\lmax, D)$ plane are shown in \Fig{fig:DvsLmax}, again using 99\% contours. Almost the entire kinematically allowed region is accessed, especially by the boundary case, which as we will see later is the most jetty of our samples.    However, if we plotted  $75\%$ and 50\% contours (not shown), much stronger correlation across all our simulations would be evident, with $D\approx \hat D$.  This suggests that for sufficiently complex signals $D$ is most useful on tails of distributions, but may add only limited information beyond $\lmax$ in the cores of distributions.

%%%%%%%%%%%%%%%%%%%%%%%%%%%%%%%%%%%%%%%%%%%%%%%%
%
%%%%%%%%%%%%%%%%%%%%%%%%%%%%%%%%%%%%%%%%%%%%%%%%

One might wonder how $D$ and event isotropy are correlated, and whether they are complementary.  They are clearly not entirely redundant:  for planar events $D=0$  while event isotropy can take a range of values, as we will see later.   It is possible that using both observables together would be valuable for identifying complex signals, and this would be interesting to study.

%%%%%%%%%%%%%%%%%%%%%%%%%%%%%%%%%%%%%%%%%%%%%%%%%%%%%%%%%%%%%%%%%%%%%%
\section{Multiplicities and Event Isotropy}
\label{sec:JM}

In models of new physics involving cascade decays, parton showers, or other nonperturbative dynamics within a hidden sector, it is common for many  hidden particles to be simultaneously produced \cite{Strassler:2006im}. 
If these SM-singlet objects then rapidly decay to Standard Model particles, the final state may have a large multiplicity of observable particles. 
Depending on the event kinematics and the jet algorithm employed, such events may also have a large multiplicity of jets.
%.
In some cases these multiplicities can themselves be used as a tool to identify signals of new physics.

In quasi-spherical events, the event isotropy scales as the square root of the particle multiplicity, as seen in \Eq{eq:theoryminimum} and \Fig{fig:healpixAvg}.
Indeed we saw this effect in some of our simulations; see \cite{paper1}.
We should therefore ask, as we investigate event isotropy, whether it is strongly correlated with particle or jet multiplicity. 
In this section we will see that the correlations between these quantities are weak.
We will also point out that  event isotropy has some intrinsic advantages as an observable that multiplicities do not share.
%As such it is an observable that can capture the richness of the hidden sector while remaining relatively insensitive to many other details that depend on the interaction between the sectors, as well as on SM dynamics.
%

 %

%%%%%%%%%%%%%%%%%%%%%%%%%%%%%%%%%%%%%
\begin{figure}[t!]
\centering
\subfloat[]{
      \includegraphics[width=0.44\textwidth]{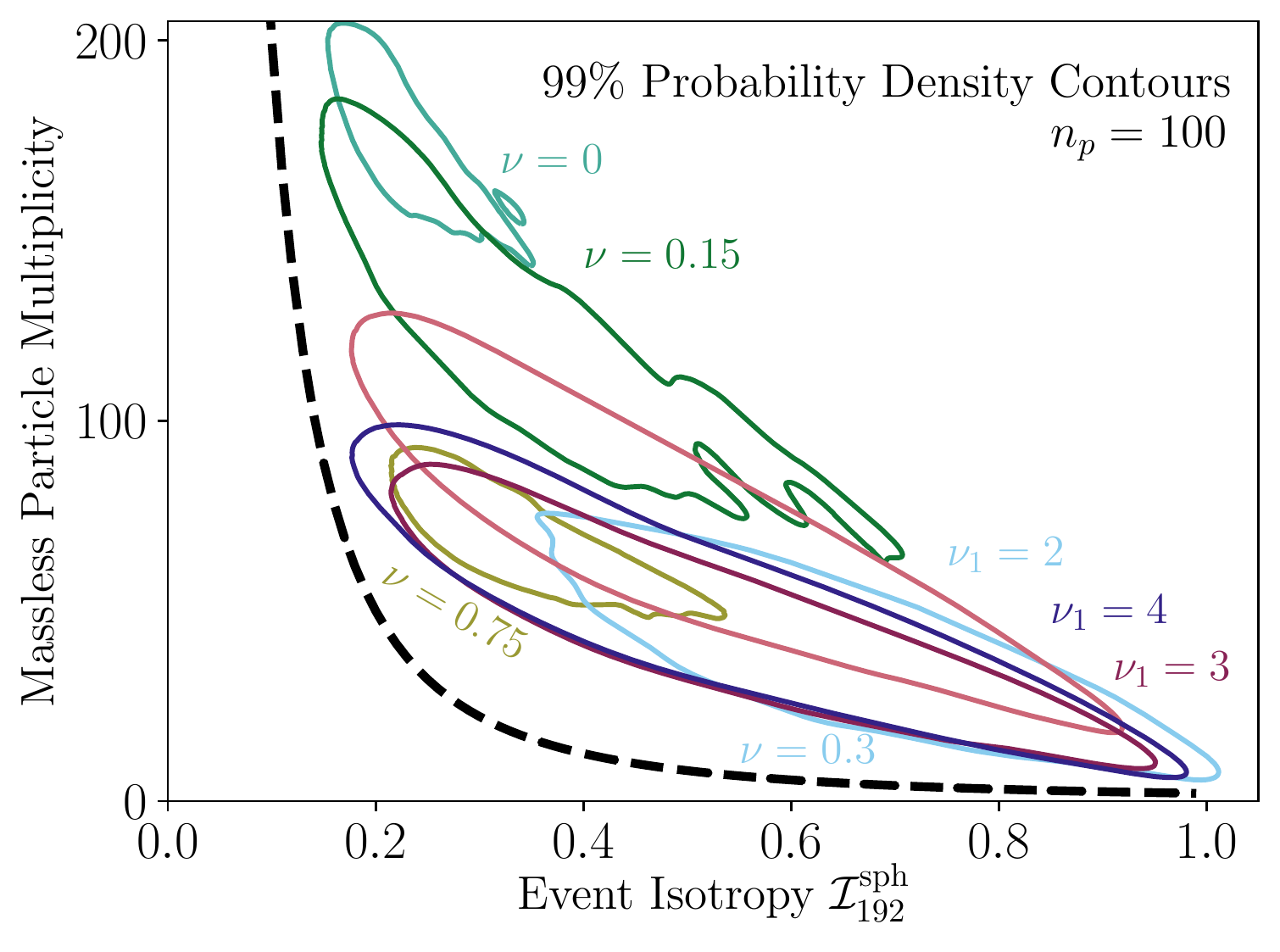}
     }
\hfill
\subfloat[]{
      \includegraphics[width=0.46\textwidth]{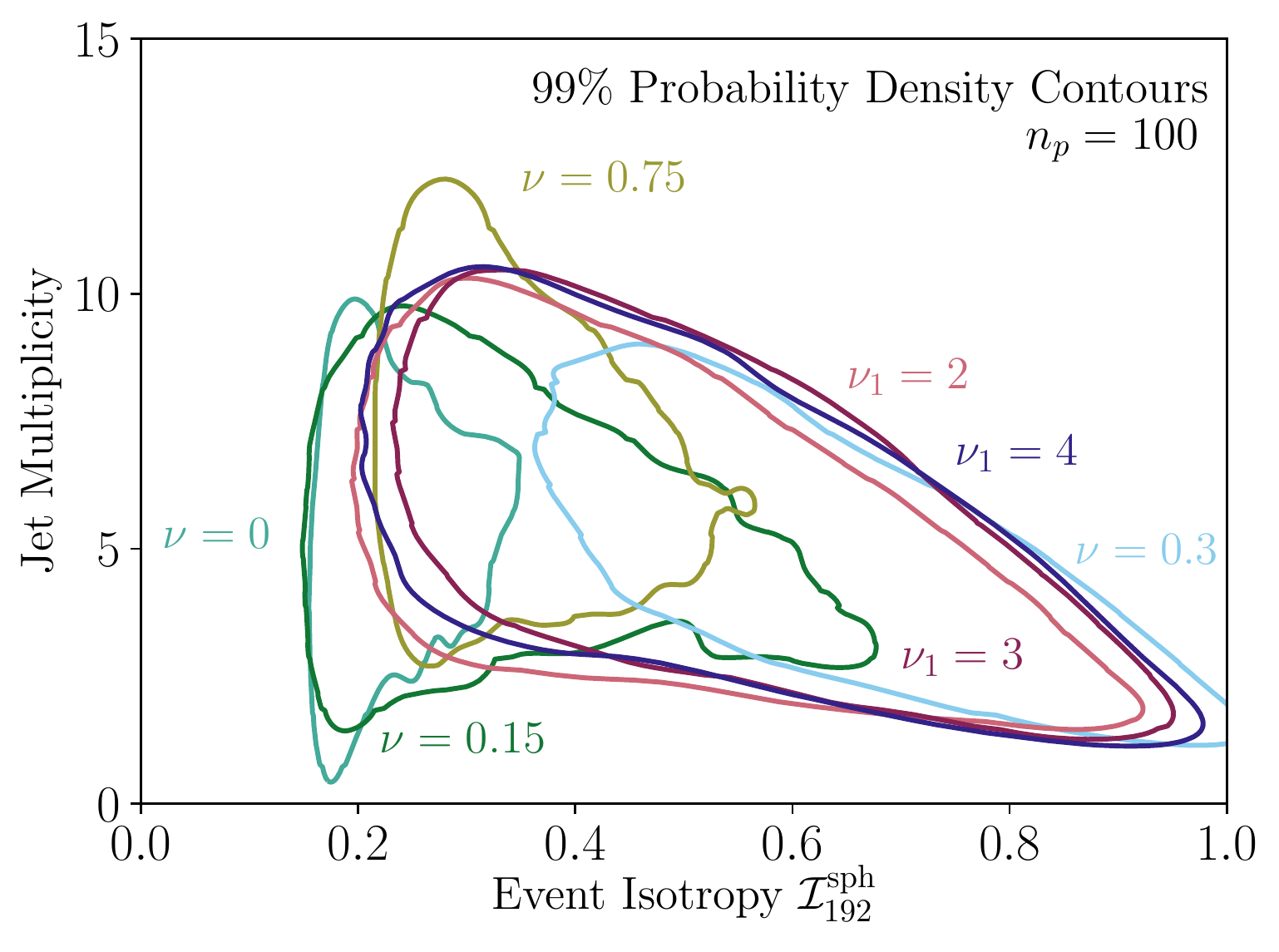}
     }
\caption{(a) Contours enclosing 99\% of the events in our simulated samples are shown in the plane of event isotropy versus ``massless particle multiplicity,'' where the massless particles in question are the decay products of the hidden stable hadrons.  Shown as a dashed line is the theoretical minimum \Eq{eq:theoryminimum}; we see that events do not occur to the left of this line, but occur broadly to its right, without a strong correlation. (b) The same, for anti-$k_T$ jets with $R=0.04$ and energy $>0.05$ of the event's energy, showing that samples with similar jet multiplicity can have drastically different event isotropy distributions.}
\label{fig:mults}
\end{figure}
%%%%%%%%%%%%%%%%%%%%%%%%%%%%%%%%%%%%%%
%

As an observable for a real detector, particle multiplicity depends on how the hidden sector particles decay to the Standard Model.
Recall that in our simulations we bypass the model-dependent issues of particle multiplicity by allowing each hidden stable hadron (HSH) to decay to two massless particles, stand-ins for elementary SM particles. 

The correlation of event isotropy with this ``massless particle multiplicity,'' which is twice the multiplicity of HSHs, but in general would not be experimentally observable, is shown in \Fig{fig:mults}a.
For each of our simulations, we plot a contour enclosing 99\% of the events within the plane of event isotropy versus massless particle  multiplicity. 
For fixed particle multiplicity $k$, the event isotropy (bounded from below and to the left by the theoretical minimum %$\sqrt{2/k}$,
  \Eq{eq:theoryminimum}, shown as the dashed curve) is broadly distributed within our various simulations. 
Since this idealized multiplicity is little correlated with event isotropy, the same would be true of any experimentally measurable definition of particle multiplicity, which would suffer the additional handicap of model-dependence.
 
For jet multiplicity, which has less model-dependence, a similar plot is shown in \Fig{fig:mults}b. 
To define our jets, we use the spherical anti-$k_T$ algorithm in FastJet with a radius of $R=0.4$ \cite{Cacciari:2011ma}.
The jets we count should be selected above a fixed energy fraction,  $f\equiv E_\text{jet}/ E_\text{total}>\efmin$; otherwise we will end up counting very soft jets that have no hard particles at their core. 
We select $\efmin$ as follows.

In a perfectly spherical event, the area and energy fraction of a typical jet is approximately $[\pi (0.4)^2]/ [4 \pi] = 0.04$.
While this would in principle allow for 25 equal area jets with energy fraction $f=0.04$, in practice the algorithm always yields some jets with smaller area. 
Empirically, when clustering a discrete spherical event in the shape of $\mathcal{U}_{3072}$ (a \texttt{HEALPix} \cite{2005ApJ...622..759G} pixelation of a sphere with $\kref=3072$ massless particles), we find 26 jets with $f$ between 2\% and 4\%, and several very soft jets with $f <1\%$. 
Choosing jets with $f\geq \efmin=0.04$ is too aggressive, since, in a quasi-spherical event with finite multiplicity and random angles, random fluctuations can cause jets to occasionally exceed 4\%. 
Meanwhile we should not choose $\efmin\gg 0.04$;  since the number of  jets is capped at $1/\efmin$,  dynamic range would be lost.  
We therefore choose $\efmin=0.05$ to maximize the dynamic range while assuring that the typical soft spherical event has zero jets above $\efmin$.

Figure \ref{fig:mults}b shows there is very limited correlation, within our samples, between event isotropy and jet multiplicity.
%This is not conceptually surprising.
This is not surprising; for instance, a pure dijet event whose two jets have $f\gg \efmin$ would have event isotropy near 1, while a spherical event with two small back-to-back lobes, and thus two jets with $f\sim \efmin$, would have event isotropy near its theoretical minimum.
We can see evidence for both extremes in the figure; the two-field and bulk-boundary simulations that give multiple narrow jets and little soft physics lie near the theoretical expectation of $\sqrt{2/n_{\rm jet}}$, while the single-field simulations that give near-spherical events have much smaller event isotropy.
More generally, as we will see below, event isotropy is affected not only by jet multiplicity but also the angular distributions of the jets, their relative energies, and the distribution of low-energy radiation in the event.

Let us note that event isotropy has intrinsic advantages over either particle or jet multiplicity.
Particle multiplicity is an event-wide measure, and can be can be useful for identifying signatures from hidden sector phenomena; see, e.g., Refs.~\cite{Strassler:2008fv,Strassler:2008bv,csaki2009ads}.
%
%But the massless particles we use in our simulations are not generically observable (unless they are photons or light leptons).
%
The measurable multiplicity of SM particles will not reflect the hidden cascade directly; it depends on the specific details of the portal(s) to the SM.
Depending on the branching fractions of the hidden particles to decay to leptons, neutrinos, photons, quarks or gluons,  wildly different numbers of final state particles may result.
There are also experimental challenges with this variable; for instance, the angular clustering among these final-state particles may also affect the efficiency with which they are identified.
Thus, although particle multiplicity may be useful in reducing SM backgrounds, it is unlikely  to reliably inform us concerning the nature of the hidden sector.
%Moreover, the radiation pattern of the cascade does not correspond to the multiplicity of observable particles.

By contrast, the kinematic distribution of energy, and therefore the multiplicity of jets found by a jet algorithm, is much less sensitive to these model-dependent details \cite{Strassler:2008fv}.
This is reminiscent of the original reason for introducing jets to study QCD events \cite{Catani:1991pm,Ellis:1991qj}, although the dynamics in this case is more general.
For example, the jet multiplicity is not very sensitive to QCD showering and hadronization of SM decay products, and is also insensitive to decays within the hidden sector of a boosted hidden hadron.
For this reason jet multiplicity, and other observables based on the kinematics, more directly reflects the physics of the hidden cascade and its initial decays to the SM than do the details of the SM particles in the final state.
In particular, a boosted hidden particle will lead to collimated energetic SM decay products that inherit its boost.
Meanwhile softer SM particles can emerge from collections of nonrelativistic HSHs, which emit back-to-back particles in a random orientation.
Thus the number of hard jets found in clustering the SM decay products, and the more isotropic softer emission underneath them, may be correlated with the initial pattern of hidden sector particles that produced them \cite{Strassler:2008fv}.

Nevertheless, the multiplicity of jets is sensitive both to the jet algorithm and to the inevitable cut on jet momentum imposed in counting the jets.  
The latter creates dependence on small kinematic effects that may move a jet above or below a cut, or split it into two.  
Event isotropy, like other event shape variables, is based solely on the energy distribution in the calorimeter, and requires neither a jet algorithm with a jet energy cut nor a counting of individual particles. 
And as we have seen, it reflects more general properties of the events than either particle or jet multiplicity.

%
%In our studies below, done in the context of our toy model, cascades begin with a heavy hidden sector hadron with KK-number $n_p \gg 1$ that decays recursively within the hidden sector down to the lightest hadrons, which are stable against decay to other hadrons (hidden stable hadrons, or HSH). 
%
%We take the HSH to decay to two massless particles, which serve as a proxy for SM particles. 
%The KK-number of the apex particle $n_p$ is a parameter of the model. 
%
%As shown in \cite{paper1}, the pattern of three-hadron couplings becomes independent of the mode number for sufficiently large KK-number \cite{paper1}, so the kinematics of the cascade and therefore formation of jets is largely independent of $n_p$.
%
%Meanwhile the massless particle multiplicity increases with $n_p$, so it is less stable than jet multiplicity.
%
%Moreover,  in a realistic model the particle multiplicity depends in detail on exactly how the massless particles of our toy model are replaced with SM particles, while jet multiplicity is generally less sensitive to this issue.

%%%%%%%%%%%%%%%%%%%%%%%%%%%%%%%%%%%%%%%%%%
\section{A Simple Estimator for Understanding Event Shape Analytics}
\label{sec:SimpleModel}
%%%%%%%%%%%%%%%%%%%%%%%%%%%%%%%%%%%%%%%%%

In this section we will obtain further insight into the behavior of event shape variables across events that interpolate between pencil dijets and quasi-spherical.  We do this by constructing families of easily understood events, for which certain event shape observables have simple analytic approximations. We focus on events that have a discrete azimuthal symmetry, defining the symmetry axis to be the $z$ axis.  

The observables considered in this discussion are $\iso{sph}{192}$ and $\lmax$. As noted in \Sec{sec:EventShape}, the advantage of $\lmax$ is that it is easily calculated, unlike thrust, and is linear in momentum fractions, unlike the $C$ and $D$ parameters. 
We noted in \Fig{fig:tvslambdamax} that $\lmax$ is strongly correlated with thrust in our classes of events. We will see in  \Sec{sec:simResults} that event isotropy is less correlated with $\lmax$; our goal in this section is to explain why.

\subsection{The Multiprong Estimator}
\begin{figure}
\centering
\subfloat[]{
       \includegraphics[trim = 45 50 75 0, clip, width=0.3\textwidth]{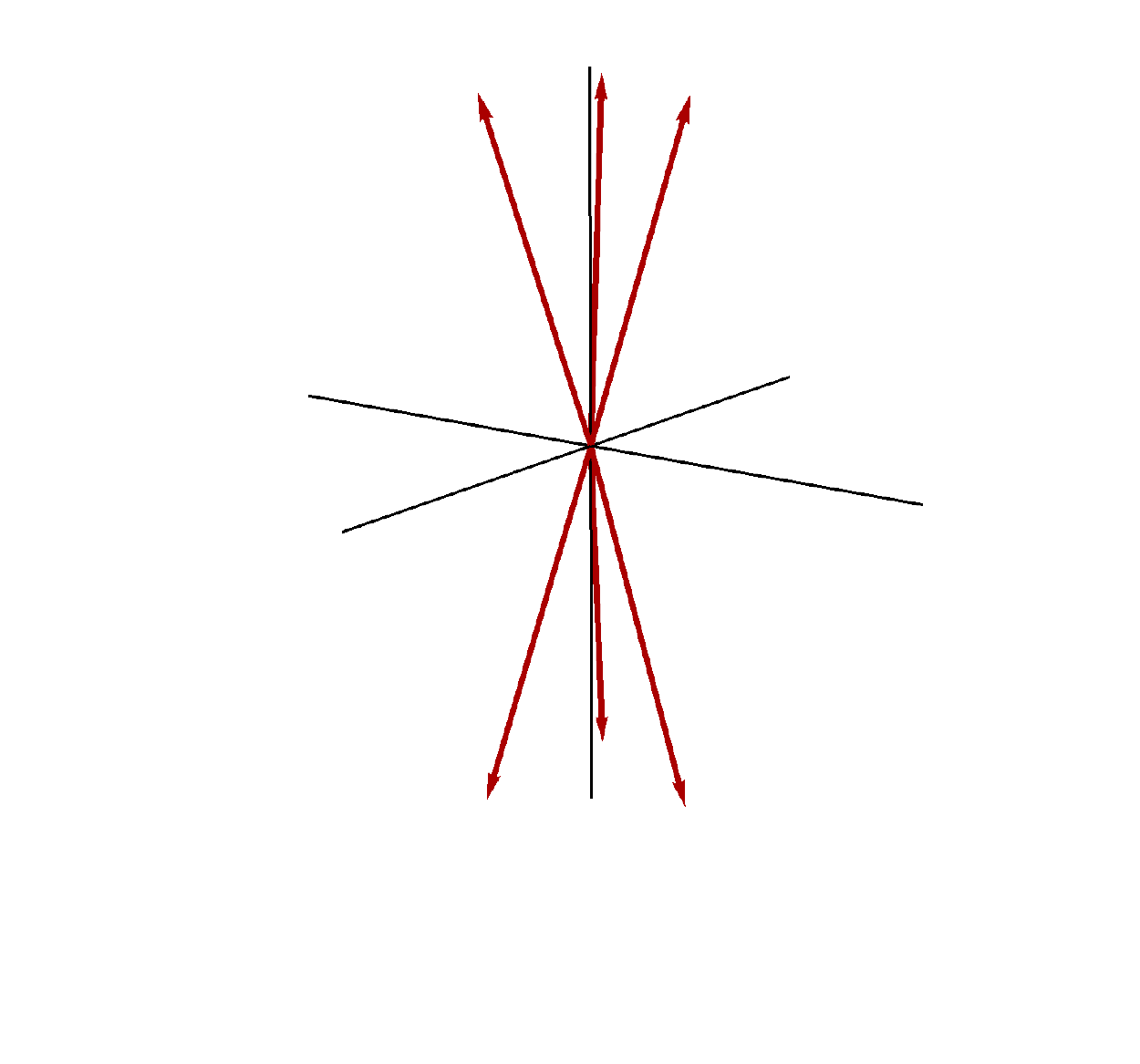}
     }
     \hfill
\subfloat[]{
       \includegraphics[trim = 45 50 75 0, clip, width=0.3\textwidth]{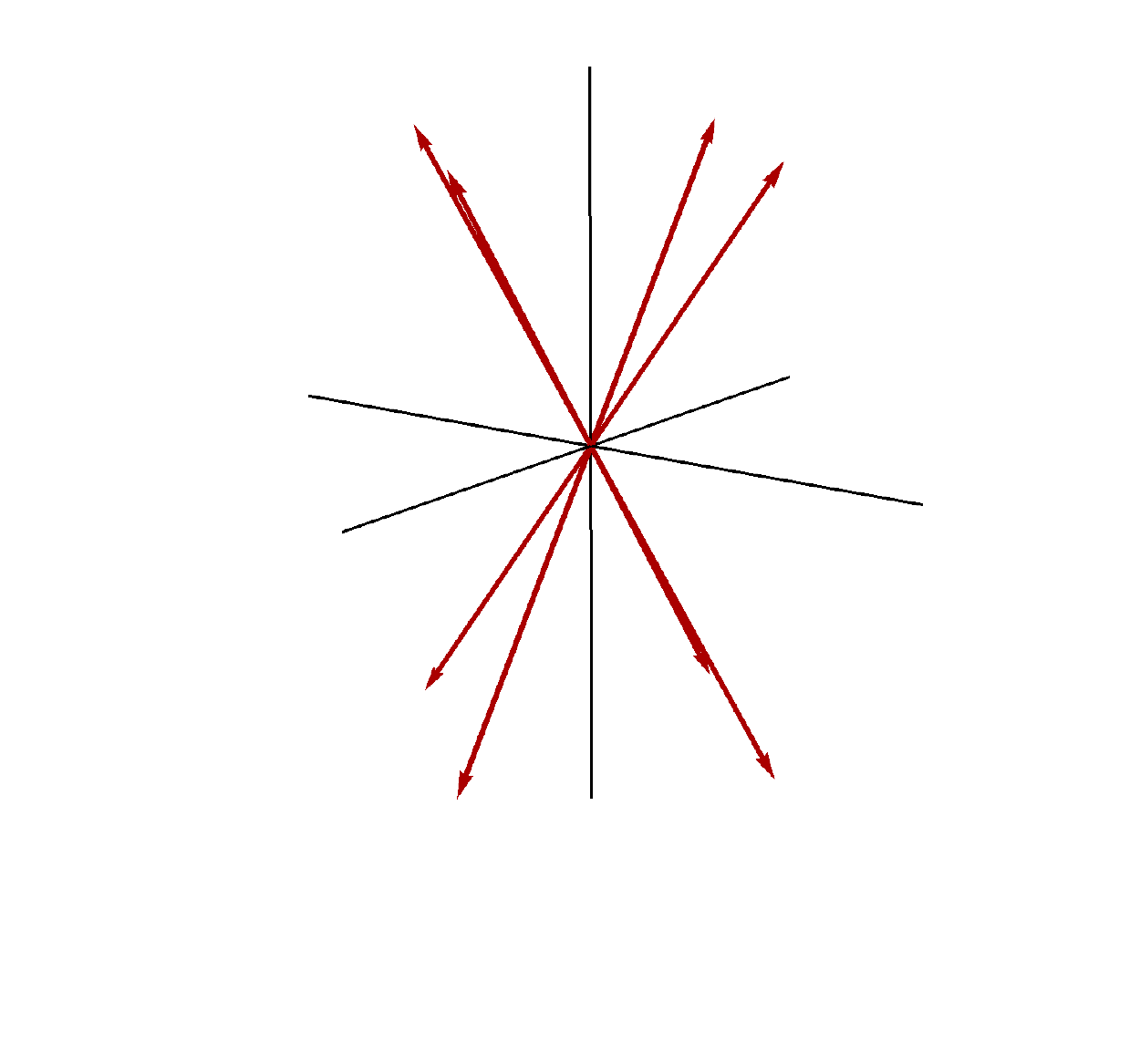}
     }
     \hfill
\subfloat[]{
       \includegraphics[trim = 45 50 75 0, clip, width=0.3\textwidth]{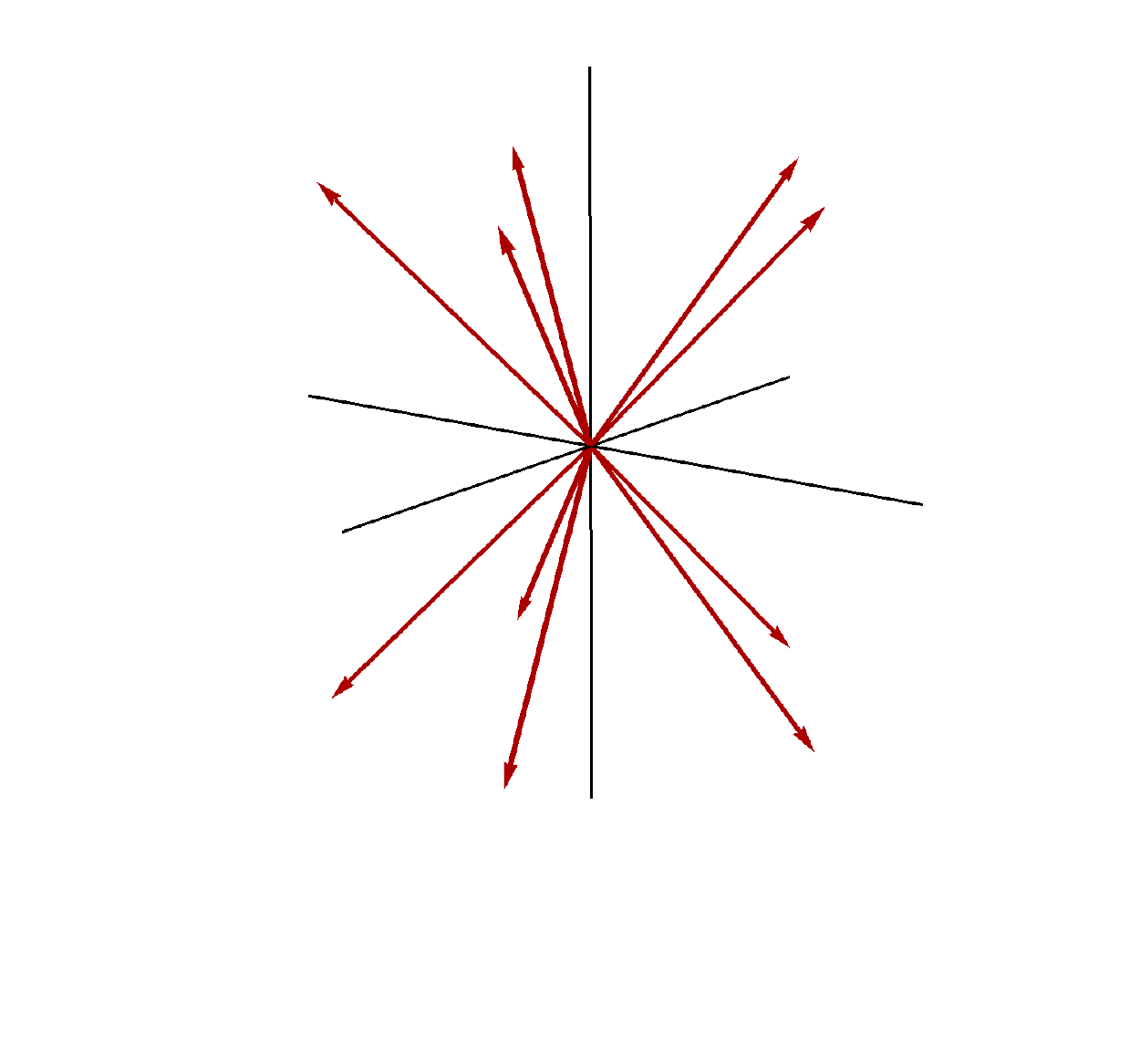}
     }
     \hfill
\caption{Visualizations of \Eq{eq:vecDef} for (a) $N=3$, $\bar{\theta} = \frac{\pi}{10}$, (b) $N=4$, $\bar{\theta} = \frac{\pi}{5}$, and (c) $N=5$, $\bar{\theta} = \frac{\pi}{4}$.}
\label{fig:simplestExamples}
\end{figure}
Consider a family of events  with $2N$ particles of momentum $p/N$, half arrayed uniformly around the positive $z$ axis and the other half around the negative axis, all at an angle $\bar\theta$ to the $z$ axis; see \Fig{fig:simplestExamples}. (Since event shape observables are dimensionless, the magnitude of the momentum is irrelevant  at this stage, but we retain it for later use.) We align the particles' momenta along unit vectors 
\be
\hat n_j \equiv 
 \begin{cases} 
      \left(\sin\bar\theta \, \cos \frac{2\pi j}{N},\sin\bar\theta  \, \sin \frac{2\pi j}{N},\cos\bar\theta  \right)&j \in \{0,1,\dots,N-1\} \\
      \\
           \left(\sin\bar\theta \, \cos \frac{2\pi j}{N},\sin\bar\theta  \, \sin \frac{2\pi j}{N}, -\cos\bar\theta  \right)&j \in \{N,N+1, \dots,2N-1\} 
   \end{cases}
\label{eq:vecDef}
\ee
such that the lower hemisphere mirrors the distribution in the upper hemisphere. 
Some examples of these configurations are visualized in \Fig{fig:simplestExamples}.

In practice the configurations that will be useful for our results in \Sec{sec:simResults} have $3\leq N\leq 5$, for which our estimator explores physically relevant configurations.  The case $N=2$ is special, the $\hat n_j$ lie in a plane; note the equations given below require modifications. For $N\gg 1$,  the vectors form a cone, atypical for events with $2N$ jets.

Because of the discrete azimuthal symmetry, the sphericity tensor defined in \Eq{eq:spherTensor} is diagonal.  We restrict ourselves, for the moment, to $\cos\bar\theta>\cos\theta_\text{crit}= 1/\sqrt{3}$, for which the tensor's largest eigenvalue is
\begin{equation}
\lmax\equiv S^{(1)}_{zz}=\cos^2\bar \theta
\label{eq:lambdaMaxSM}
\end{equation}
with remaining eigenvalues $S^{(1)}_{xx}=S^{(1)}_{yy}=\half\sin^2\bar \theta$. 
The critical angle indicates where the $\lmax$ eigenvector jumps from the $z$ axis to the $xy$-plane. 
\begin{figure}
\centering
\includegraphics[trim = 75 80 75 80, clip, width=0.42\textwidth]{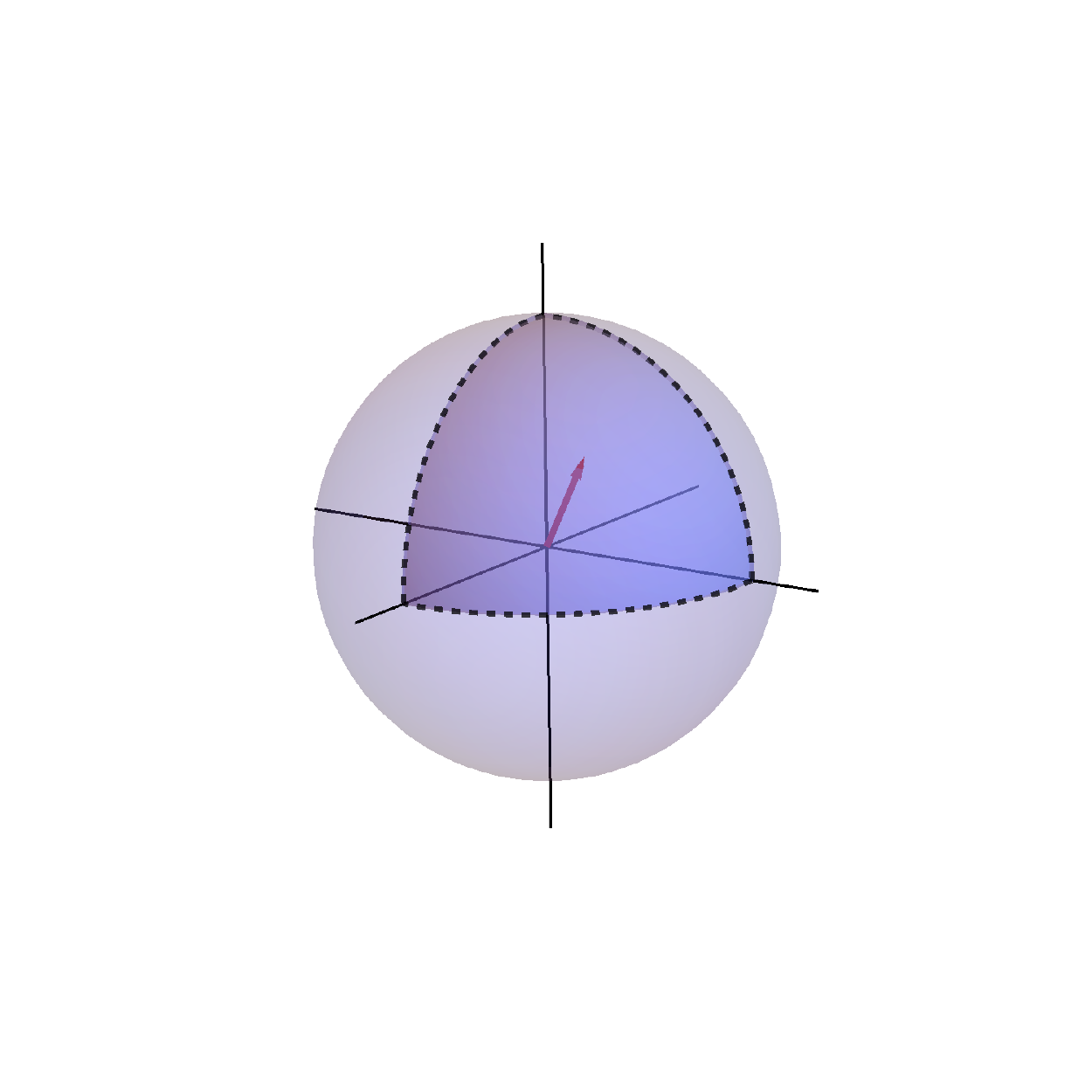}
\caption{An illustration of how the energy is spread from a multiprong event into a sphere. 
 For  general $N$ the energy is spread across the surface of a half-wedge occupying $1/N$ of a hemisphere.  For $N=4$, the special case shown, each half-wedge is a equilateral spherical right triangle. }.
\label{fig:slice}
\end{figure}

Let us now estimate the event isotropy for these events.
For these analytic estimates, we assume a perfectly isotropic reference sphere ($\kref=\infty$) except where explicitly noted.
By symmetry, the energy of each vector must be spread across a slice of the upper or lower hemisphere of azimuthal width $2\pi/N$ (\Fig{fig:slice}). The isotropy is therefore:
\be\label{eq:Isotropy2N}
{\iso{sph}{\infty}} =  \frac{2N}{4\pi} \int_0^1 {\rm d}\cos\theta \int_{-\frac{\pi}{N}}^{\frac{\pi}{N}}  {\rm d}\phi \
\frac{3}{2}\sqrt{1 - \hat n_0\cdot \hat r}
= \frac{3N}{\pi} \int_0^1 {\rm d}\cos\theta \ \sqrt{1 - \hat n_0\cdot \hat r_0} \ E\left(\frac{\pi}{2N},-\frac{2 \sin \theta \sin \bar \theta}{1 - \hat n_0\cdot \hat r_0}\right),
\ee
where $\hat n_0$ is defined in \Eq{eq:vecDef}; $\hat r$ is a unit vector parameterized by the  angles $\theta, \phi$; $\hat r_0$ is $\hat r$ with $\phi=0$; and $E(\phi,m)$ is the incomplete elliptic integral of the second kind, defined as 
\begin{equation}
E \left(\phi, m \right) = \int_0^\phi \mathrm{d}\theta \left(1-m \sin^2\theta \right)^{1/2}.
\end{equation}

In practice, we compute event isotropy using a discretized reference sphere with $\kref=192$ pixels. The effect of finite $\kref$ on \Eq{eq:Isotropy2N} is very small for these events, whose small multiplicity makes their EMD insensitive to the details of the reference sphere. 

The integral in \Eq{eq:Isotropy2N} has no closed form.  For large $N$ and small $\bar\theta$---the near dijet regime, where $\lmax\approx 1$ ---the approximation simplifies to
\be
{\iso{sph}{\infty}}% \approx 1 - \left(\sqrt2-\frac12\right)\bar\theta 
\approx 1 - 0.91\left(1-\frac{\pi^2}{6}\frac{1}{N^2}\right)\sqrt{1-\lmax} + \dots \ .
\ee
This decreases linearly in $\sqrt{1-\lmax}\sim \bar\theta$ (recall \Eq{eq:lambdaMaxSM}), with a slope whose magnitude decreases  as $N$ increases. 
 While we will see this qualitatively in \Fig{fig:earlyCasc} and \Fig{fig:lateCasc},  the expansion in $1/N^2$ converges slowly and the approximation is quantitatively poor.

In the opposite regime, where $\lmax$ is near $1/3$, simple considerations offer an estimate. As noted in \Eq{eq:theoryminimum}, an event with $k$ isotropically distributed particles on a sphere has $\iso{sph}{\infty} \sim \sqrt{2/k}$, and by symmetry it will have $\lmax\sim 1/3$.   For $N=2$ or $N\gg 1$ the radiation pattern is never isotropic, but for $N\sim 3-5$ the distribution of vectors is fairly regular when $\lmax=1/3$.  We therefore expect
\be
{\iso{sph}{\infty}}\left(\lmax=\frac{1}{3}\right) \approx \frac{1}{\sqrt{N}} = \{0.577,\ 0.500,  \ 0.447\} , \quad (N=3,4,5) \ .
\ee
This compares favorably with the exact result in \Eq{eq:Isotropy2N}, which gives\footnote{The integral \Eq{eq:Isotropy2N} can be organized in another way for $\lmax=1/3$ and $N=4$, using the symmetries of the problem.  In this maximally symmetric situation, each of the eight vectors sits at the center point of a right-angle equilateral spherical triangle with 1/8 the area of the sphere.   In this case one rotates one of the particles onto the $z$ axis, and dividing the surrounding equilateral triangle into three isosceles triangles, one can show 
\be
{\iso{sph}{\infty}} = \frac{6}{\pi} 
\int_{-\frac{\pi}{3}}^{\frac{\pi}{3}}  {\rm d}\phi \  
\left(1 - \frac{\cos(\phi)}{\sqrt{\frac{1}{2} +  \cos^2(\phi)}}\right)^{3/2}
 \approx 0.513.
\ee
} $\{0.596,\ 0.513,\ 0.466\}$.
As a comparison, numerical calculation with $\kref=192$, averaged over random orientations, gives $\langle \iso{sph}{192}\rangle = 0.517$ for $N=4$.

Note that the sphericity tensor and therefore $\lmax$ have no functional dependence on $N$, whereas event isotropy is sensitive to the multiplicity. This already shows, in a trivial way, that event isotropy can vary for a fixed sphericity tensor --- one may simply vary $N$ holding $\bar{\theta}$ fixed. 

In the special case of $N=2$, we must modify our approach when calculating $\lambda_\text{max}$. 
Since the momenta of the event are constrained to a plane, only two eigenvalues are nonzero.
The leading eigenvalue is still given by \Eq{eq:lambdaMaxSM}, but the remaining eigenvalue $\lambda_\perp$ is now 
\begin{equation}
\lambda_\perp = 1 - \lambda_\text{max} = \sin^2 \bar{\theta}.
\end{equation}
The maximum value of $\bar{\theta}$ before the $\lmax$ eigenvector jumps is determined by
\begin{equation}
\cos \theta_\text{crit}^{N=2} = \frac{1}{\sqrt{2}}.
\end{equation}
When using $N=2$, we calculate the observables  \Eq{eq:lambdaMaxSM} and \Eq{eq:Isotropy2N} over this restricted range in $\bar{\theta}$. 

Though we will not need it directly in this paper, in other contexts it will be useful to extend the range of the estimator, by adding a dijet on the $z$ axis to these events. Specifically, to our $2N$ vectors $\vec p_n$, $n=1\dots 2N$ defined in \Eq{eq:vecDef}, we can add vectors $q\hat z$ and $-q\hat z$ ($q>0$). The sphericity tensor remains diagonal but the maximum eigenvalue is sensitive to the additional dijet. From \Eq{eq:spherTensor},
\be
\lmax = \max\left(\frac{q + p\cos^2\bar\theta}{q+p}, \frac{p  \sin^2 \bar\theta}{2(q+p)}\right).
\label{eq:lambdaMPD}
\ee
The critical angle beyond which the $\lmax$ eigenvector no longer points in the $z$ direction is determined by
\be
\cos \theta_\text{crit} = \sqrt{\frac{1}{3}  - \frac{2q}{3p}}.
   \label{eq:thetacritgeneral}
\ee

 In App.~\ref{app:AddDijet} we show that a reasonable approximation to the event isotropy, subject to certain constraints on $N$, $\bar\theta$, and $q/p$, is 
\bea\label{eq:withpolarcap}
\iso{sph}{\infty}
&\approx&\left(\frac{q}{p+q}\right)^{3/2} + 
\frac{3N}{\pi} \int_0^{p/(p+q)} {\rm d}\cos\theta \ \sqrt{1 - \hat n_0\cdot \hat r_0} \ E\left(\frac{\pi}{2N},-\frac{2 \sin \theta \sin \bar \theta}{1 - \hat n_0\cdot \hat r_0}\right)
 \ ,
\eea
where $\hat n_0, \hat r, \hat r_0$ are defined as in \Eq{eq:Isotropy2N}. %The approximation holds subject to certain constraints on $N$, $\bar\theta$, and $q/p$, described in App.~\ref{app:AddDijet}.  
Expansion in $q/p$ shows the event isotropy is a non-monotonic function.  Specifically, for fixed $N\sim 3-5$ and fixed $\lmax\sim 1/3-1/2$, varying $q/p$ can cause event isotropy to vary by  $\sim \pm0.05$ compared to its  value at $q/p=0$.  

\subsection{Combining this model with a soft sphere}
 \label{sec:multiprongplussphere}
 
Up to this point we have considered only events that model jetty physics.  Now we extend the estimator to events that have a small number of harder prongs embedded in a soft spherically symmetric background.  
%This turns out to be reasonably simple because we are using \ms{the?} the proper metric in computing isotropy, in contrast to \cite{Cesarotti:2020hwb}.  

Take one of the multiprong events that we have discussed above (with parameters $N, p, q,\bar \theta$) and rescale its particles' energies so that its total energy is $x\leq 1$. Next consider a ``soft sphere'' --- a perfectly uniform spherical distribution of $k=\infty$ soft particles --- with total energy $(1-x)$. Combining these two components gives an event with energy 1, whose sphericity tensor is a linear combination of the tensors of its two components.  Because the sphericity tensor of a sphere  is $1/3$ times the identity matrix, the combined event's sphericity tensor is diagonalized by the same rotation matrix for any $x$.  Meanwhile, as long as $\kref=\infty$, it costs no energy to move the soft sphere to the reference sphere, while the particles in the multiprong event will have their rescaled energy distributed across the reference sphere exactly\footnote{Importantly, this is only true if the metric $d_{ij}$ that defines the Energy Mover's Distance satisfies the triangle inequality; this is true of our definition in this paper but not true of the one used in \cite{Cesarotti:2020hwb}.  Otherwise it may be preferable to move the energy in the soft sphere.}
 as for $x=1$. 
It follows that
%In this case the $S^{(1)}$ tensor and event isotropy  are both simple linear combinations of the two.
%Fix $p$, $q$ and $\bar\theta$; rescale the momenta so that $p+q=x$.  
%Since a sphere contributes $1/3$ to each $S^{(1)}$ eigenvalue, 
\be \label{eq:lambdaaddsphere}
\lmax(x) = x  
\lmax(x=1) + \frac{1-x}{3} \nonumber,
\ee
\be\label{eq:addsphere}
\iso{sph}{\infty}(x) = %x\iso{sph}{}(x=1)+ (1-x)\iso{sph}{}(\text{sphere}) = 
x\iso{sph}{\infty}(x=1) \ .
\ee
%because $\iso{sph}{}(\text{sphere})=0$ when both the event and the reference sphere of comparison are uniform, infinite multiplicity spheres. 

Thus, if we were to fix $q/p,N,\bar \theta$ in \Eq{eq:addsphere},  $\lmax$ and $\iso{sph}{\infty}$ would be perfectly correlated as a function of $x$. But event isotropy may be varied, while holding $\lmax$ and $x$ fixed, by adjusting these other parameters. Examples are shown in %\msi \Sec{sec:simResults} and\msf 
\App{app:AddDijet}.

In realistic events and practical computations, the number of particles $k$ in the soft sphere is finite, and the number of pixels $\kref$ in the reference sphere is finite as well.  The effect on $\lmax$ is minimal, but the event isotropy of a given event will more significantly depend on the orientation of the soft sphere relative to the reference sphere.  If $k=\kref$ the two spheres could be perfectly aligned, in which case there is no EMD associated to the soft sphere and the result in \Eq{eq:addsphere} holds with $\infty$ replaced by $\kref$.  For any other relative orientation, the event isotropy is larger; the average over orientations is given by \Eq{eq:approx}.  On the other hand, the event isotropy is bounded from above by the convexity property \Eq{eq:EMDconvexity}.  
In a set of events with random orientations,  the average 
$\iso{sph}{\kref}(x)$ thus lies somewhere between these extremes:
 \be\label{eq:addsphererange}
x\iso{sph}{\kref}(x=1) \leq \iso{sph}{\kref}(x) \leq  x\iso{sph}{\kref}(x=1)+ (1-x)\sqrt{2 /{\rm min}(k, \kref)}\ . 
\ee 
Since the square root is at most of order $0.1$ for $k\gtrsim\kref\sim192$, this equation predicts $\iso{sph}{192}(x)$ to within a range of at most $0.1(1-x)$.
Further discussion of the $x$-dependence and other details  is given in \App{ap:EMD}.

%%%%%%%%%%%%%%%%%%%%%%%%%%%%%%%%%%%%%%%%%
%%%%%%%%%%%%%%%%%%%%%%%%%%%%%%%%%%%%%%%%%
\section{Simulation Results}
\label{sec:simResults}
%%%%%%%%%%%%%%%%%%%%%%%%%%%%%%%%%%%%%%%%%

In this section, we present results from our simulations; see \Sec{sec:intro} for a brief review and \cite{paper1} for further details. We establish that the event isotropy $\iso{sph}{}$ measures different properties of the events  than do certain traditional event shape variables. In particular, $\iso{sph}{}$ is not as strongly correlated with $\lmax$ as is  thrust; see \Fig{fig:tvslambdamax}(b), and the simulated events occupy substantial regions in the ($\iso{sph}{}, \lmax)$ plane. 
%even when we control for the number of jets (e.g., selecting a fixed number of jets above 5\% of the total event energy, a cut that we motivated in \Sec{subsec:JMeventshape}). 
%Thus, the information captured by event isotropy is distinct from that which is captured by traditional ways of characterizing events. %Furthermore, we will see that the range over which our model events vary in the $({\cal I}, \lambda_{\rm max})$ plane is close to that captured by the simple estimator discussed in \Sec{sec:SimpleModel}.

%%%%%%%%%%%%%%%%%%%%%%%%%%%%%%%%%%%%%%%%%%%%%%%%
\subsection{General Considerations}
%%%%%%%%%%%%%%%%%%%%%%%%%%%%%%%%%%%%%%%%%%%%%%%%
\begin{figure}[!h]
\centering
\includegraphics[width=0.65\textwidth]{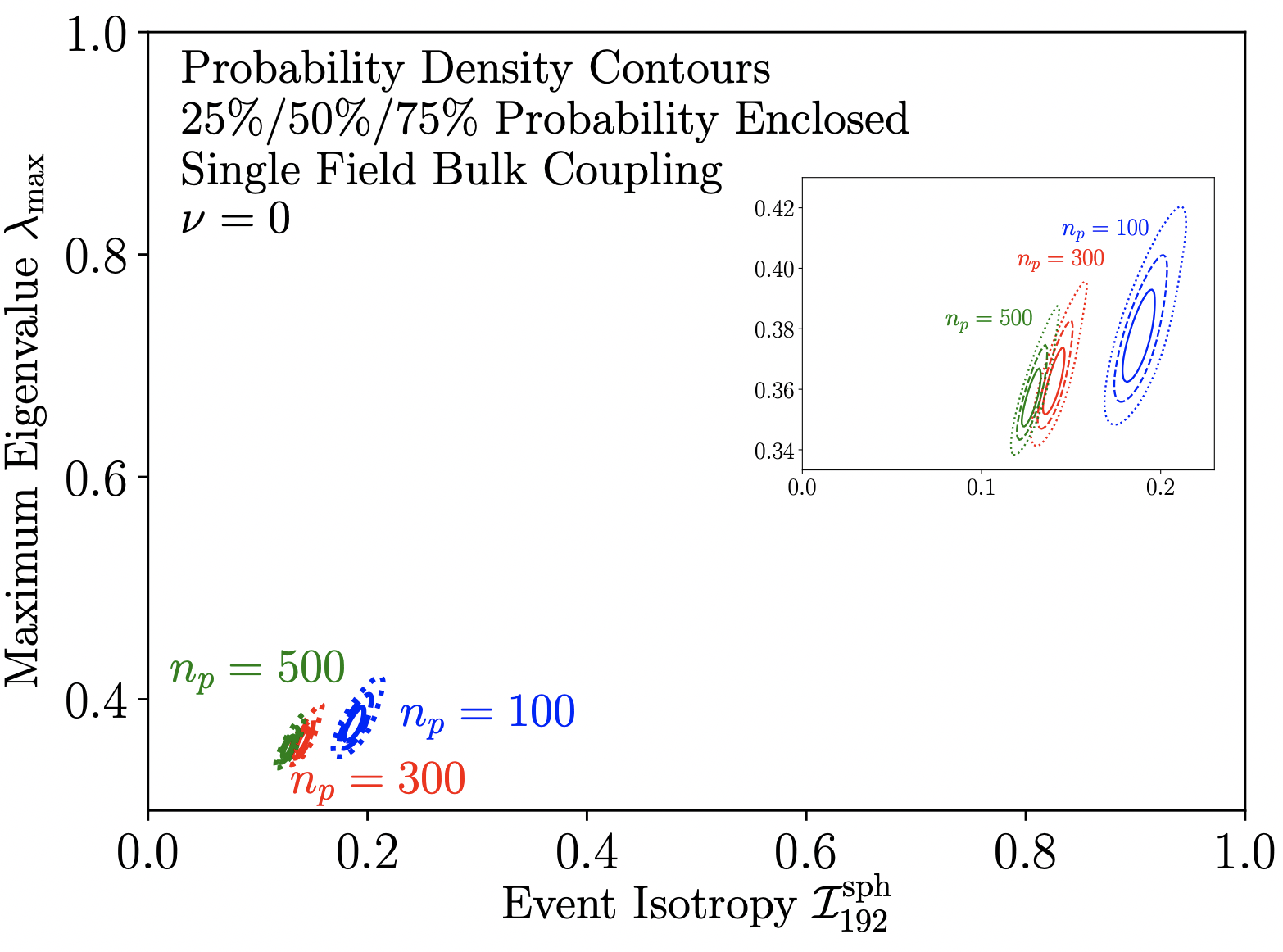}
\caption{Contours of probability density quartiles (25\%, 50\%, and 75\% containment as solid, dashed, and dotted lines) for final states in the single-field case with $\nu=0$, with varying initial parent KK mode number $n_p$.% This plot reinforces the conclusion that $\iso{sph}{192}$ captures significantly different information about the event shape than $\lmax$.  
}
\label{fig:singleFieldNoJC}
\end{figure}
%
%%%%%%%%%%%%%%%%%%%%%%%%%%%%%%%%%%%%%%%%%%%%%%%%

%%%%%%%%%%%%%%%%%%%%%%%%%%%%%%%%%%%%%%%%%%%%%%%%
\begin{figure}[!h]
\centering
\includegraphics[width=0.65\textwidth]{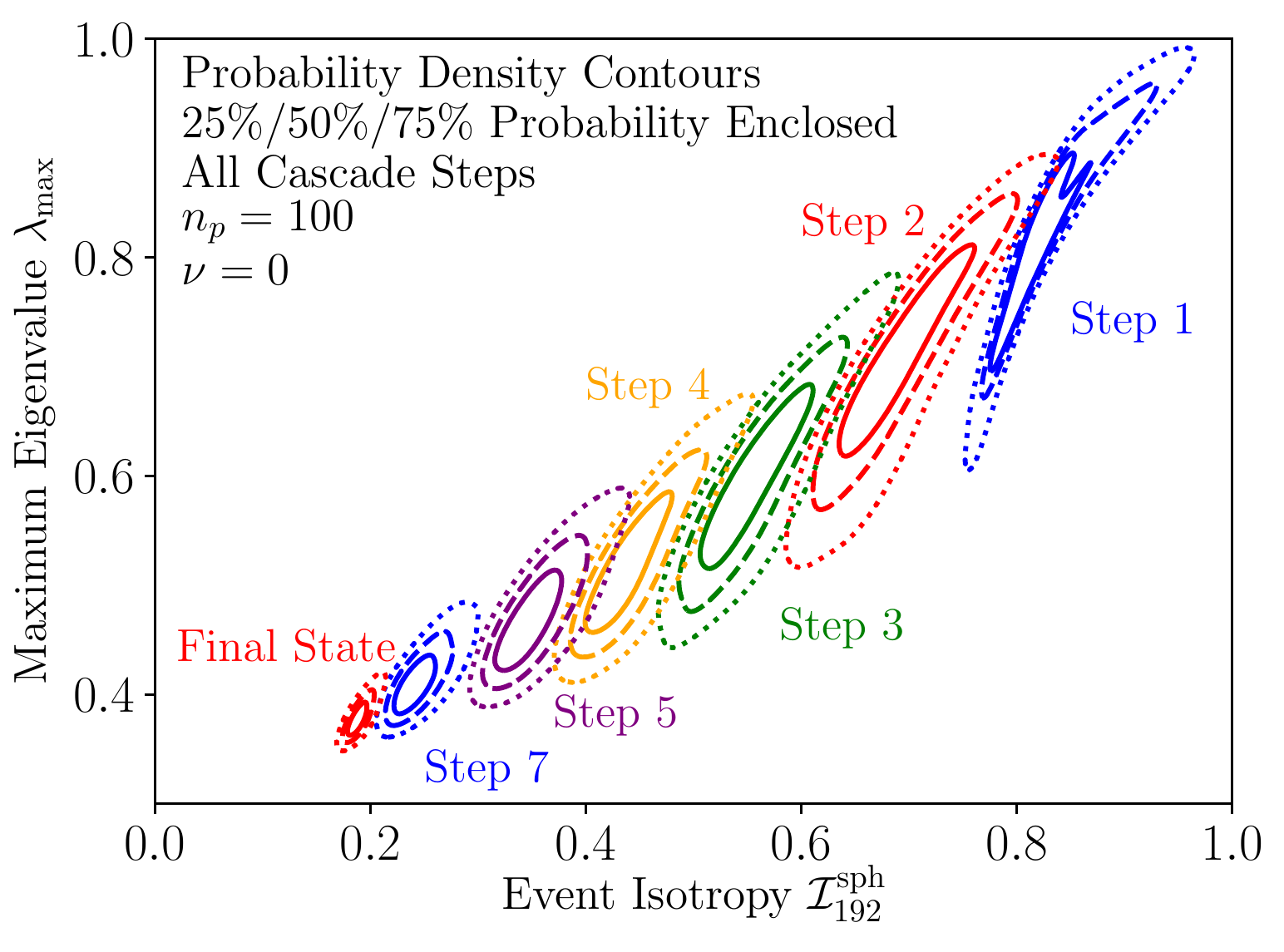}\
\caption{Contours of probability density quartiles (25\%, 50\%, and 75\% containment as solid, dashed, and dotted lines) for events stopped after a fixed number of steps in the decay cascade. We study the decay of the 100th Kaluza-Klein mode for a single scalar field with $\nu = 0$, which produces nearly spherical events \cite{paper1}. The interrupted cascades are initially jetty (upper right) and end up near-spherical (lower left).
}
\label{fig:fullNu0Cascade}
\end{figure}
%
%%%%%%%%%%%%%%%%%%%%%%%%%%%%%%%%%%%%%%%%%%%%%%%%

We begin with samples obtained from the single-field case.  A single  scalar field with a bulk $\Phi^3$ coupling and 5d mass $M$ produces a tower of interacting scalar 4d states, which we refer to as hidden hadrons.  These are labeled by a quantum number $n$, which we refer to loosely as KK-number.  A hadron with $n=n_p$ is set at rest and allowed to decay, in a cascade, to hidden stable hadrons (HSHs).  Each of these then in turn decays to a pair of massless particles, which serve as a proxy for elementary Standard Model particles.  

We first consider a field at the Breitenlohner-Freedman limit; from \Eq{eq:nuFromMass} this has $\nu=0$. In this case, the only HSH is  the state with $n=1$.   Were KK-number conserved, the hadron with $n=n_p$ would decay to $n_p$ HSHs with $n=1$, and $2n_p$ massless particles would result.  Moreover, because $m_n\sim 1.31  m_1 \times n$ , the HSHs would all be non-relativistic.  Their massless decay products would emerge with random angles but comparable energies, leading to events that tend to be highly isotropic.

As discussed in \cite{paper1}, KK-number conservation is broken, but for $\nu=0$ the breaking is small and observables are very similar to the KK-conserving case.  In particular, the average multiplicity of massless particles lies close to but slightly below 2$n_p$.  The  theoretical minimum \Eq{eq:theoryminimum} for highly spherical events with  $\gtrsim 192$ particles is $\iso{sph}{192} \sim \sqrt{2/192} \approx 0.1$.  As shown in \Fig{fig:singleFieldNoJC}, for $n_p=500$ this limit  is reached, along with $\lmax=1/3$, which applies to any distribution with triaxial symmetry.\footnote{A more detailed discussion of the dependence of event isotropy on $n_p$ may be found in Appendix B of \cite{paper1}.}  For $n_p=100$, which we will be using below, angular fluctuations within individual events make the events imperfectly isotropic, and so one finds instead a narrow distribution of $\iso{sph}{192}$ near 0.18.  The angular fluctuations also affect  the sphericity tensor, so $\lmax$ lies slightly above the theoretical lower bound of $1/3$.   Note that all contours shown here, and in most ensuing plots, indicate the first three quartiles of probability density (25\%, 50\%, 75\%), not standard deviations.   

To gain some intuition as to how physical processes can populate the $({\cal I}^{\rm sph}_{192},\lmax)$ plane,  we will consider how these variables evolve as the cascade proceeds.  We imagine the cascade as a series of steps; in each step, all unstable hidden hadrons from the previous step are allowed to decay to two others.  Thus, in the initial steps the number of hadrons after step $j$ grows as $2^j$, though when $j\sim \log_2 n_p$, this number saturates. We then imagine forcing all the hadrons after step $j$ to decay to two massless particles, from which we then construct event shape observables.  The question is what we can learn from the evolution with $j$,  illustrated in Fig.~\ref{fig:fullNu0Cascade} for $n_p=100$.  

As we just discussed, at the end of the cascade (after $11$ steps, typically), the events are near-spherical and have small  ${\cal I}^{\rm sph}_{192}$ and $\lmax\sim 1/3$, at the lower left corner. At step $j=0$, we have only the initial hadron with $n_p=100$; its decay to two back-to-back massless particles creates events that all sit at ${\cal I}^{\rm sph}_{192}=\lmax=1$, the upper right corner.  These then set the boundary conditions for the evolution, which, as the figure shows, saturates after about eight steps. We do not plot individual steps between the seventh and the final state as the contours are nearly indistinguishable.  Importantly, the evolution proceeds on a broad,  curved swath, with  much of the data lying well below a straight path connecting the initial state and final state distributions.  Comparing this with the extremely narrow correlation between $\lmax$ and thrust seen in \Fig{fig:tvslambdamax}, we see already that event isotropy provides new information.

Quantitative insight into the evolution can be obtained by using the estimator of \Sec{sec:SimpleModel}, along with facts about the jets in the event, which we will separate into ``hard'' and ``soft,'' where a hard jet has $>5\%$ of the total energy of the event.  

\begin{figure}
\subfloat[]{
\includegraphics[width=0.49\textwidth]{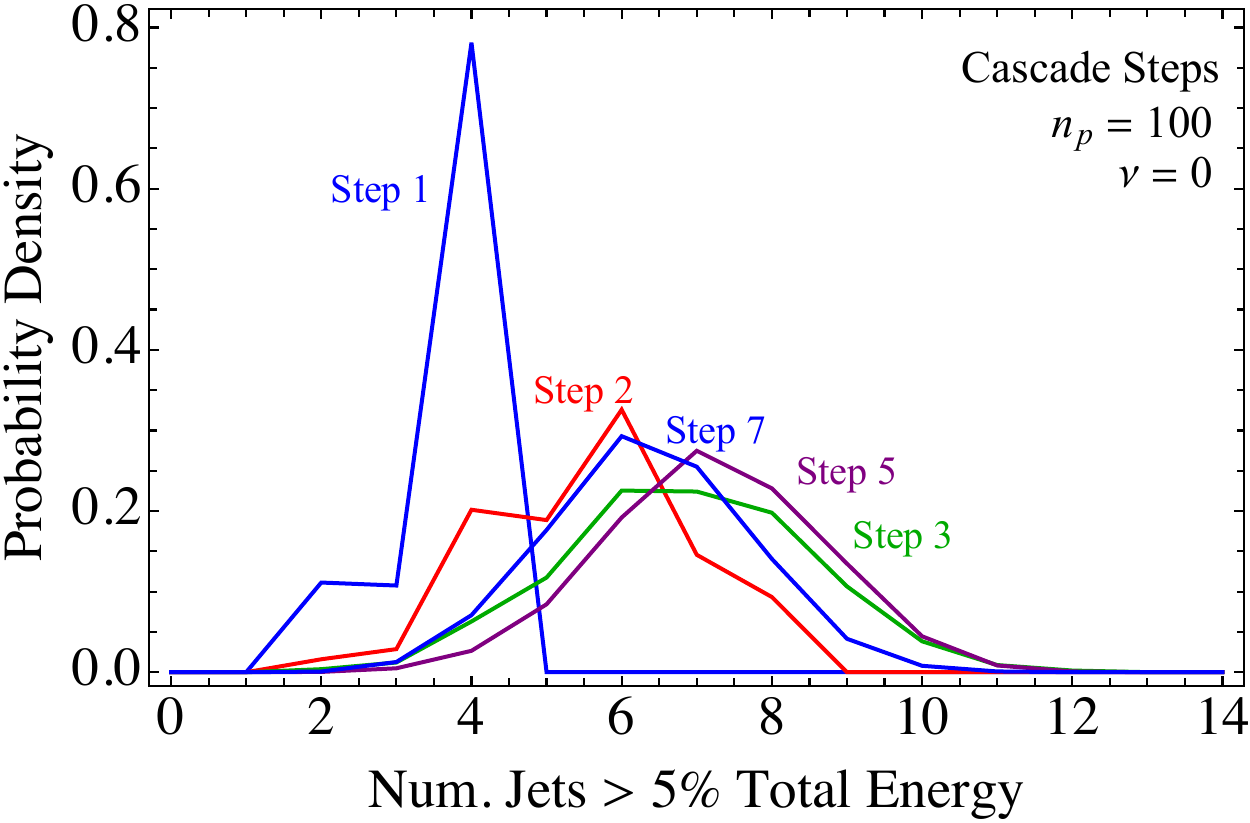} }
\hfill
\subfloat[]{
\includegraphics[width=0.49\textwidth]{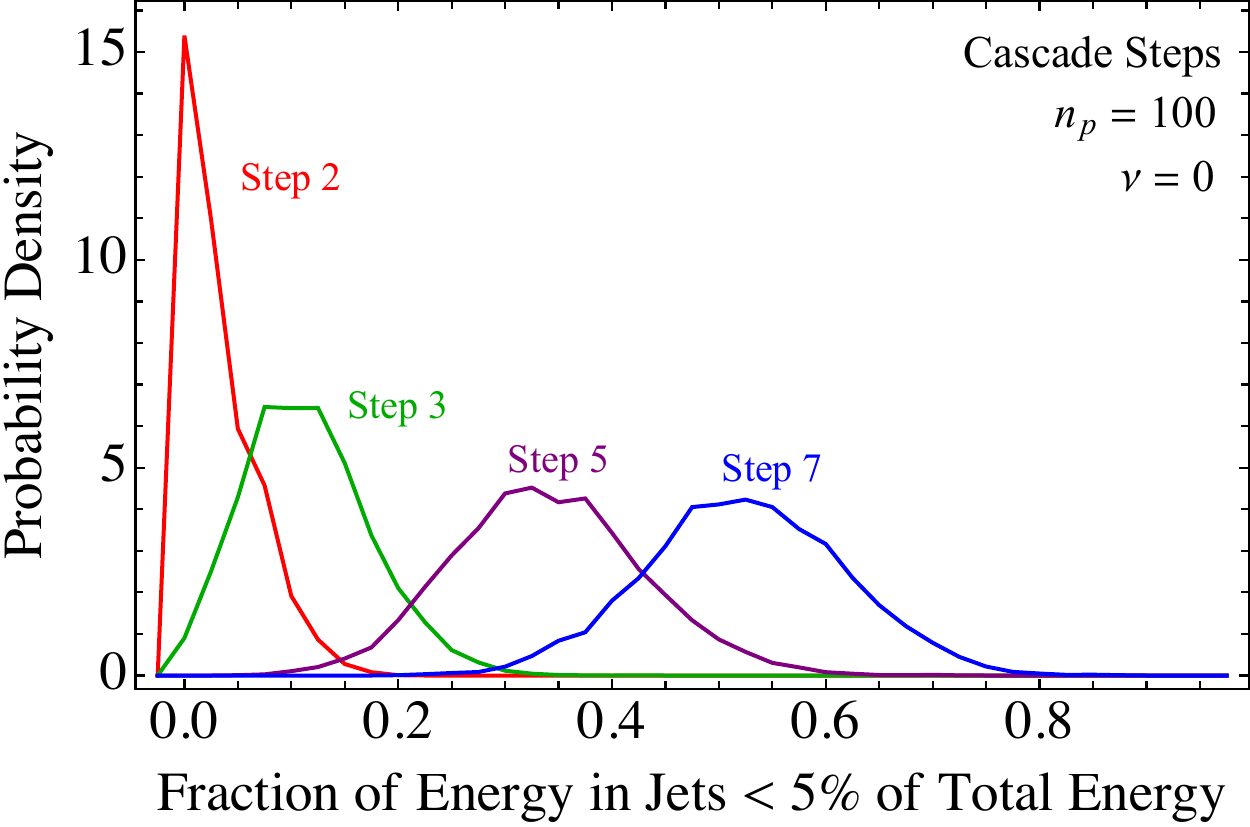}
}
\caption{Characteristics of the $\nu=0$, $n_p=100$ cascade steps shown in  \Fig{fig:fullNu0Cascade}. We present (a) the distributions of the number of hard jets for all steps shown in \Fig{fig:earlyCasc} and \Fig{fig:lateCasc} as well as (b) the fraction of the total energy outside of hard jets. (Hard jets have energy $>5\%$ of the total energy.) We see that as the cascade develops, the average number of hard jets remains near 6, but soft radiation accounts for a greater fraction of the total energy.}
\label{fig:curveLogic}
\end{figure}

\begin{figure}[!h]
\centering
\subfloat[]{
\includegraphics[width=0.33\textwidth]{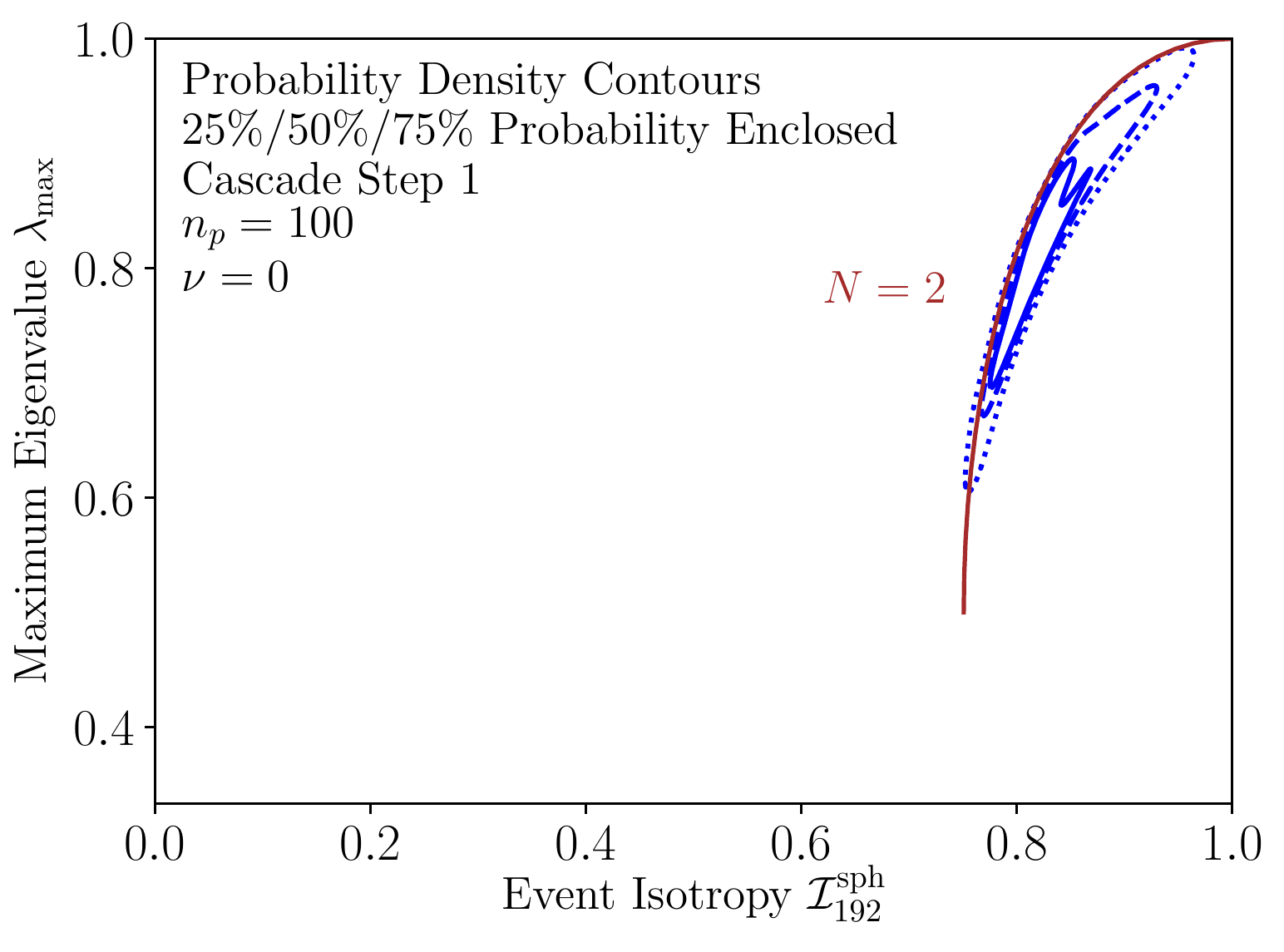}}
\subfloat[]{
\includegraphics[width=0.33\textwidth]{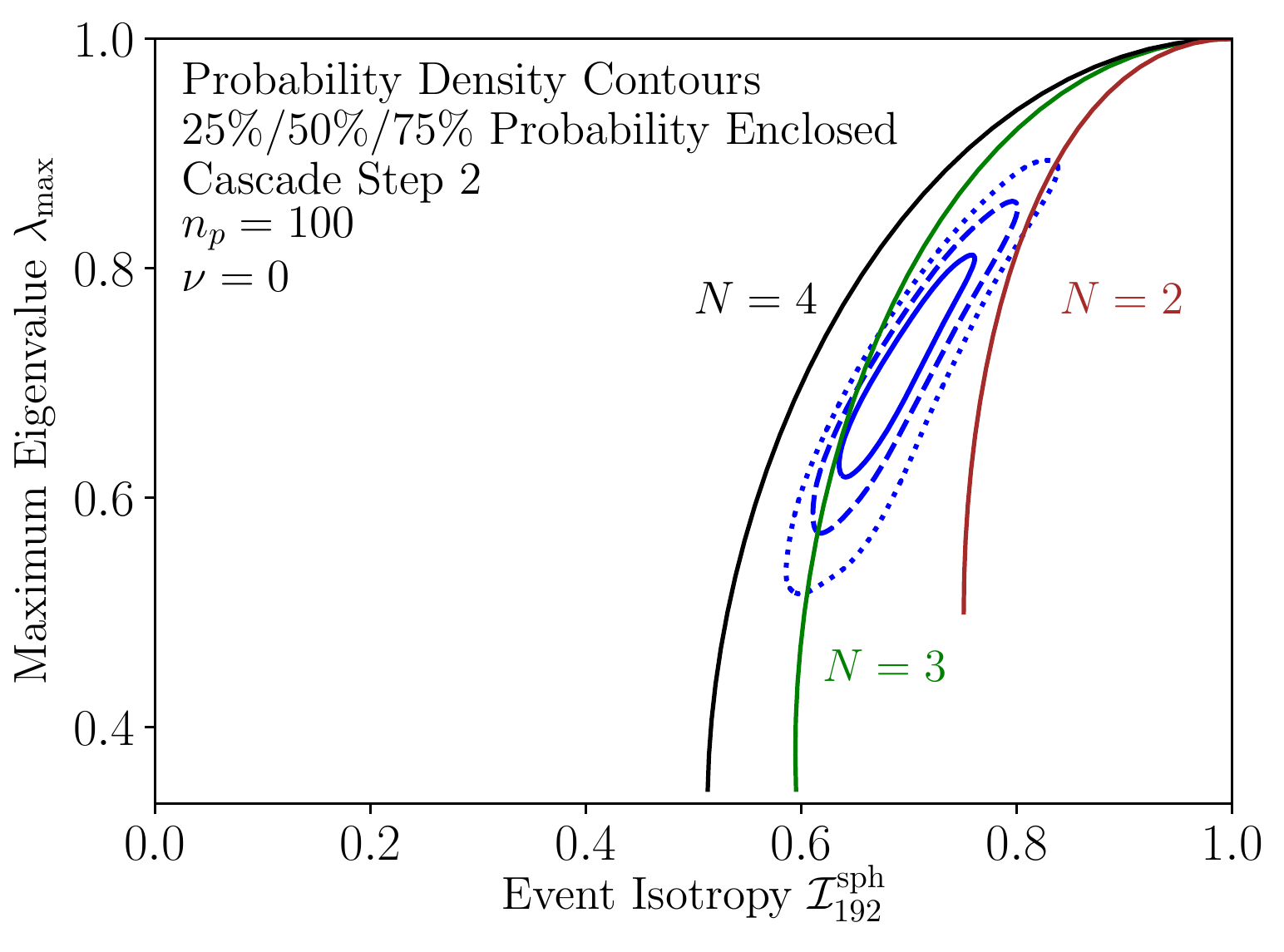}}
\subfloat[]{
\includegraphics[width=0.33\textwidth]{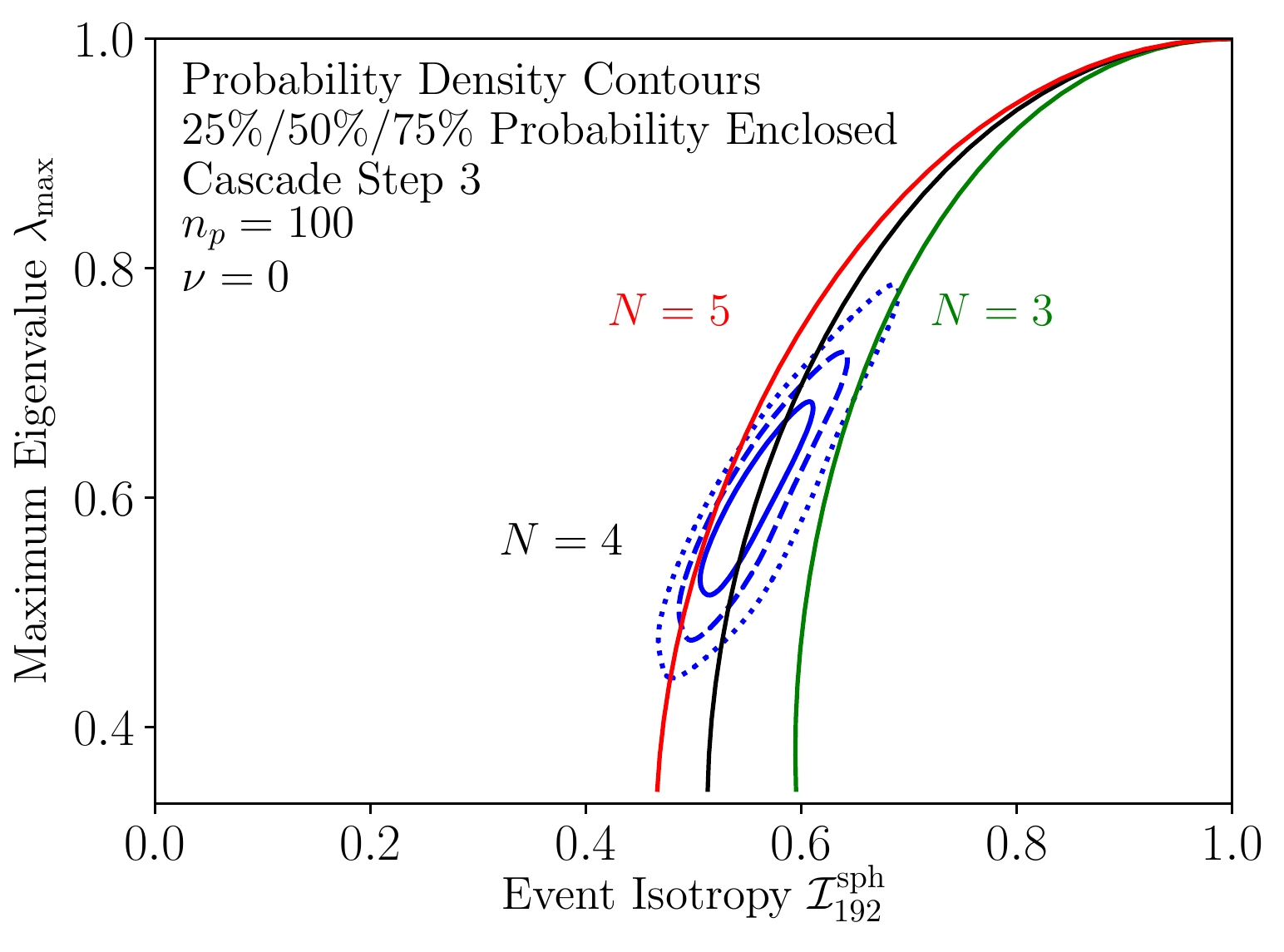}}
\caption{The quartile probability contours from \Fig{fig:fullNu0Cascade} are replotted for steps  (a) 1, (b)  2, and (c) 3, and  overlaid with curves defined by varying $\bar \theta$ in \Eq{eq:lambdaMaxSM} and \Eq{eq:Isotropy2N} for the indicated values of $N$ (with $q=0$).}
\label{fig:earlyCasc}
\end{figure}

The contours of the first three steps are replotted in  \Fig{fig:earlyCasc}, along with estimator curves from \eqref{eq:Isotropy2N}.  The values of $N$ for the curves shown  are chosen at or near one-half the value of the number of hard jets, whose distribution is shown in \Fig{fig:curveLogic}.  The curves are parameterized by $\cos\bar\theta$ over the domain  $1> \cos\bar\theta >\cos\theta_\text{crit} = 1/\sqrt{3}$, except for $N=2$ where $\cos\theta_\text{crit} = 1/\sqrt{2}$.

At the first step, the initiating hadron usually decays near threshold to two non-relativistic hadrons whose masses $m$, $m'$ may be different.  Per our procedure, the hadrons then decay to massless particles, two with energy $\sim m/2$ and two with energy $\sim m'/2$. This gives four isolated particles that are nearly contained in a plane spanned by two axes, one from the decay of each hadron, with the axes of the pairs separated by an angle $\zeta$. If the energies were all equal, this would match the multiprong estimator for $N=2, \bar \theta=\zeta/2$, giving a wide distribution in $\lmax$ whose average would be at $\lmax= 3/4$. The effect of unequal energies is generally to increase both $\lmax$ and event isotropy; this follows from the fact that when $m\gg m'$, one pair of particles dominates the event, driving it back toward the upper-right corner.  Thus the distribution is bounded at the left side by the $N=2$ estimator curve, peaks along that curve in a broad range around $\lmax=3/4$, and extends from that region mainly up and to the right.

After step 2, there are four pairs of particles in the final state. Jets are typically made from a single particle, but because of their variable energies, the likelihood of eight hard jets is small; on average there are only six hard jets, as shown in \Fig{fig:curveLogic}. It is therefore unsurprising that the events lie between $N=2$ and $N=4$ estimator curves, and are centered more on $N=3$, shifted up and right slightly by fluctuations in angles and energies.  
Although the estimator curves reach the upper right corner,  the contours have moved away it. Because our outer contour is a {75\%} containment line, it fails to reveal  a thin tail of events extending to the corner.  This tail is sparse because it is rare that four pairs of particles align (or that three of the four are soft) so as to create a near-dijet final state with $\lmax\sim 1$; typically $\lmax$ is substantially smaller.

After step 3, the distinction between soft and hard particles becomes important. Although the number of particles is usually 16, the number of particles with $>5\%$ of the total energy is typically between 5 to 12, with an average of 8.  Meanwhile, as seen in \Fig{fig:curveLogic}, the number of hard jets ranges from 4 to 10 with an average of 7.  Now, as we can see in \Fig{fig:earlyCasc}, $N=3$ and $N=5$ curves bound the events, with $N=4$ at the heart of the distribution.   Again, the random orientations of the particles disfavor $\lmax$ near one, so the contours lie away from the upper right corner. Moreover, since an average of 10\% of the energy is now in soft jets (see \Fig{fig:curveLogic}), we might also consider using \eqref{eq:lambdaaddsphere} with $x=0.9$, which would shift the estimate slightly further toward the lower left.  We neglect to do so explicitly here because with so few soft particles the approximation of a ``soft sphere'' used in \eqref{eq:lambdaaddsphere} is not yet accurate.

\begin{figure}[!h]
\centering
\subfloat[]{
\includegraphics[width=0.495\textwidth]{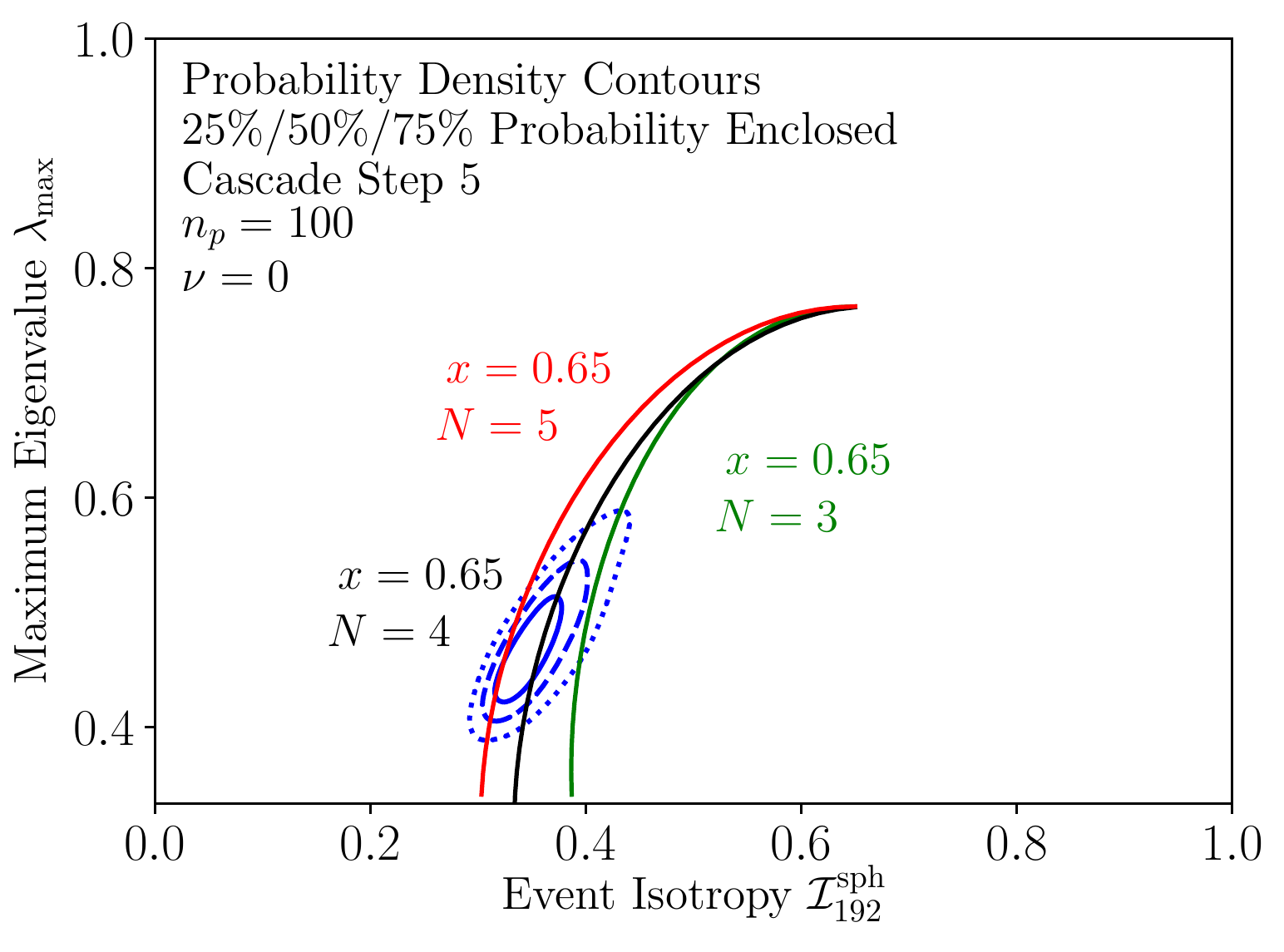}}
\subfloat[]{
\includegraphics[width=0.495\textwidth]{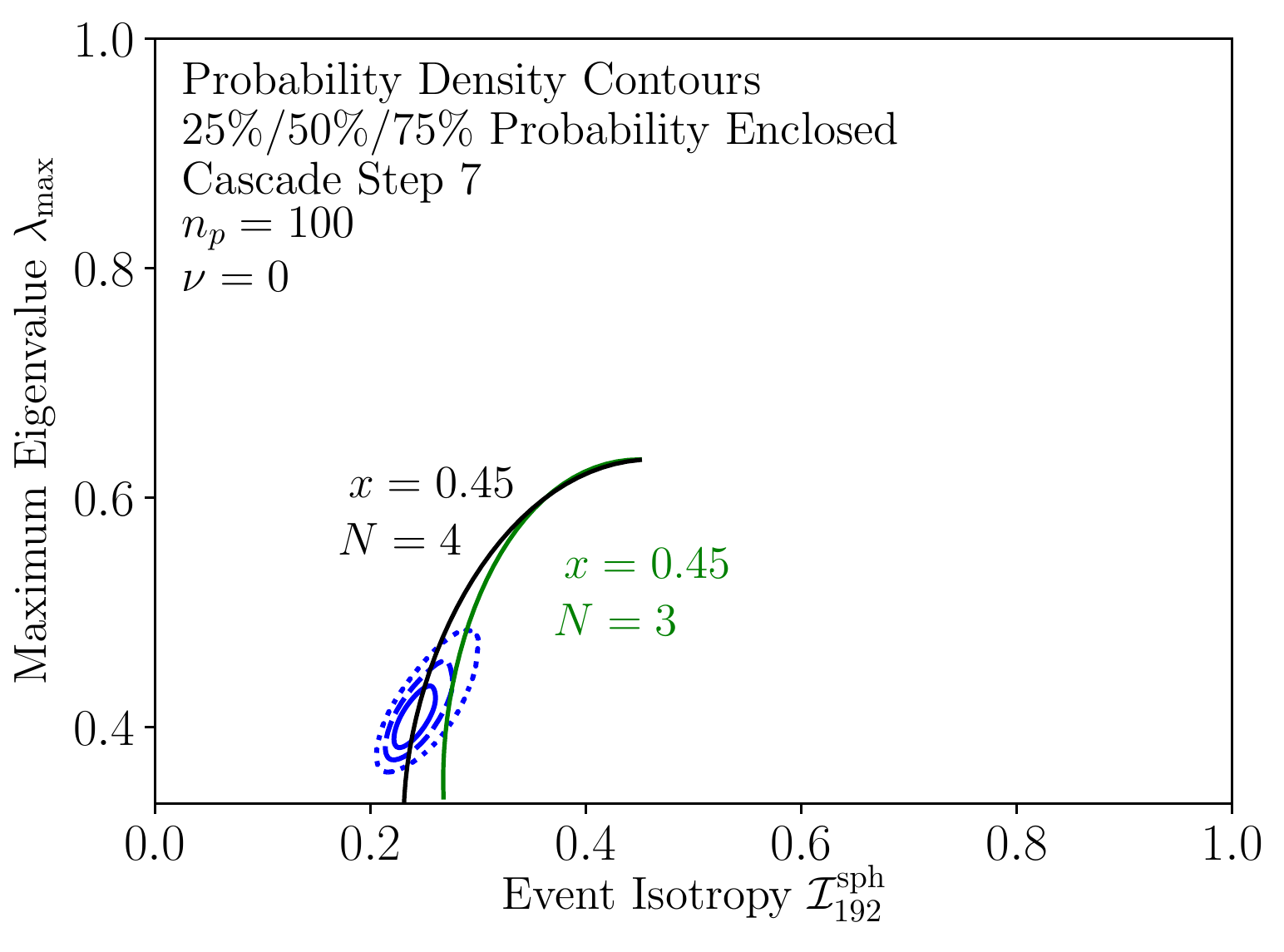}}
\caption{The quartile probability contours from \Fig{fig:fullNu0Cascade} are replotted for (a) step 5 and (b) step 7, and  overlaid with curves for varying $\bar \theta$ and indicated values of $N$ and $x$ (with $q=0$) from the multiprong estimator defined by Eqs.~(\ref{eq:lambdaMaxSM}) and,  (\ref{eq:Isotropy2N}) and (\ref{eq:lambdaaddsphere}. 
Recall $1-x$ is the fraction of total energy in the soft sphere; the soft energy in each sample is shown in \Fig{fig:curveLogic}. 
We do not show the $N =5$ curve in (b) as 10 particles with  $>5\%$ of the total energy is inconsistent with $x=0.45$.  
}
\label{fig:lateCasc}
\end{figure}

In the later steps shown  in \Fig{fig:lateCasc}, clustering of particles into jets becomes important. Also, the ``soft sphere'' becomes a better approximation, so the estimator is applied using \eqref{eq:lambdaaddsphere}.   We again choose the $N$ values of the estimator curves as one half of the average values of the jet multiplicity above 5\% of the energy. To model the soft sphere, we set $1-x$ as the average value of the fraction of total energy contained in jets with less than 5\% of the total energy, i.e., soft jets. The distributions are shown in \Fig{fig:curveLogic}. After step 5 the number of particles is now of order 30 but the number of hard jets is nearly the same as after step 3.  Meanwhile, the amount of energy in soft jets has a broad distribution averaging 35\%, so $x\sim 0.65$ on average. This shifts the estimator curves far from the upper right corner.  At the point where the curves now intersect,  the events would take the form of a dijet plus a soft sphere, but with so many particles that regime is statistically disfavored.  Instead,  the distribution is shifted to the lower part of the estimator curve, where the particles are more isotropically distributed, and $\lmax\sim 1/3$.

These same effects are seen again after step 7, only more prominently.  By this point more than 50\% of the energy is in the soft sphere, the number of hard jets has actually decreased slightly compared to step 5, and $\lmax$ lies even closer to 1/3.  With only $\sim$45\% of the energy in hard jets, the maximum number of hard jets is $\sim 9$, so we do not plot the $N=5$ curve.  The curves shown use \eqref{eq:lambdaaddsphere}and thus slightly underestimate the true event isotropy, but the effect is smaller than allowed by the inequality \eqref{eq:addsphererange}, as one can see in \Fig{fig:eviso_xdep} from \App{ap:EMD}.
%%%%%%%%%%%%%%%%%%%%%%%%%%%%%%%%%%%%%%%%%%%%%%%%
%
%%%%%%%%%%%%%%%%%%%%%%%%%%%%%%%%%%%%%%%%%%%%%%%%
%%%%%%%%%%%%%%%%%%%%%%%%%%%%%%%%%%%%%%%%%%%%%%%%
%
%%%%%%%%%%%%%%%%%%%%%%%%%%%%%%%%%%%%%%%%%%%%%%%%

%%%%%%%%%%%%%%%%%%%%%%%%%%%%%%%%%%%%%%%%%%%%%%%%
\begin{figure}[!h]
\centering
\includegraphics[width=0.65\textwidth]{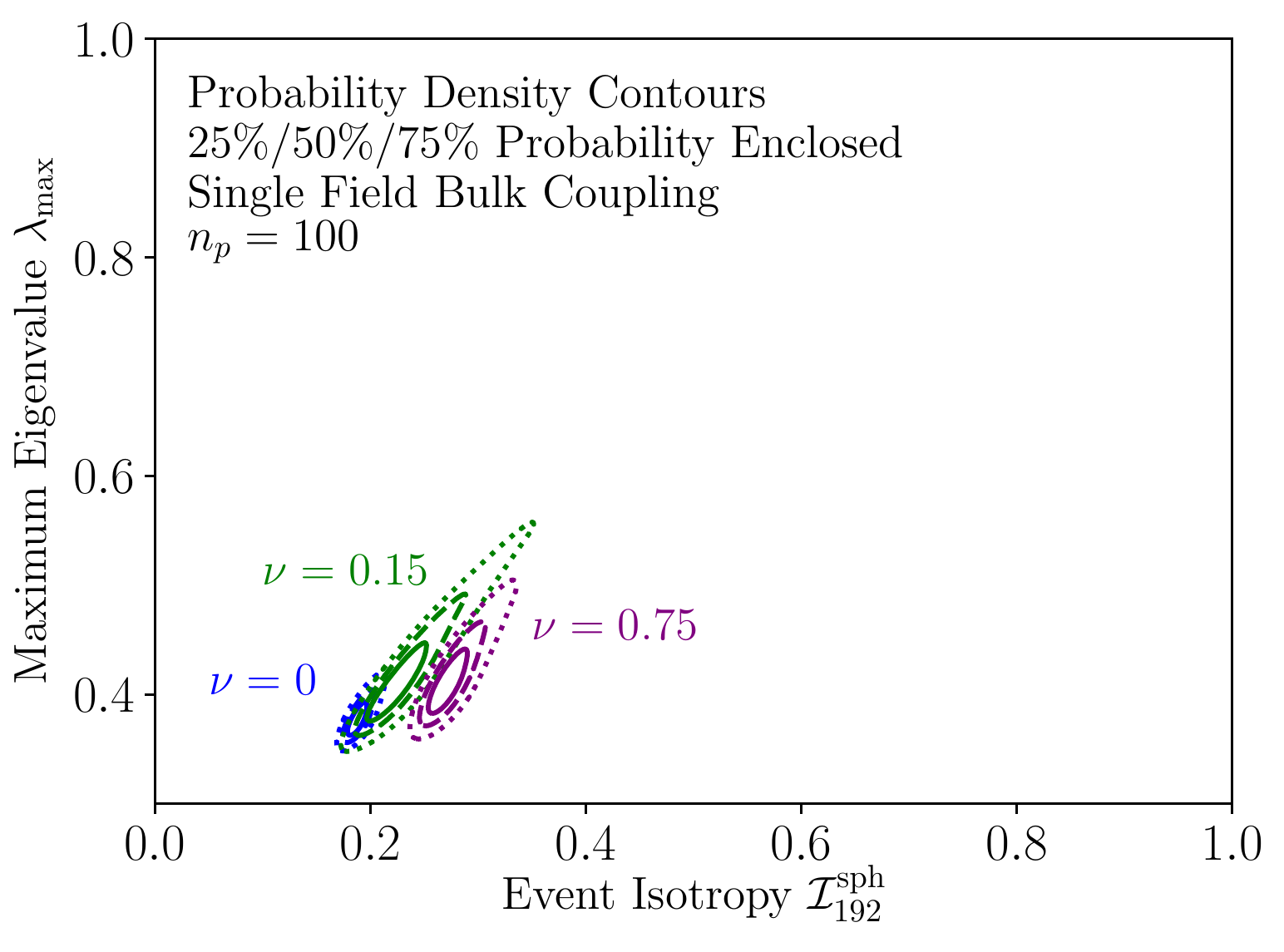}
\caption{Contours of probability density quartiles (25\%, 50\%, and 75\% containment as solid, dashed, and dotted lines) for final states in the single-field scenario, with varying bulk mass parameter $\nu$.
}
\label{fig:singleFieldj6c5}
\end{figure}
%
%%%%%%%%%%%%%%%%%%%%%%%%%%%%%%%%%%%%%%%%%%%%%%%%

%%%%%%%%%%%%%%%%%%%%%%%%%%%%%%%%%%%%%%%%%%%%%%%%
\subsection{Single field ($\nu_1 = \nu_2 = \nu_3$)}
%%%%%%%%%%%%%%%%%%%%%%%%%%%%%%%%%%%%%%%%%%%%%%%%

In Fig.~\ref{fig:fullNu0Cascade}, we considered a single field, fixing  both the bulk mass parameter ($\nu = 0$) and the initial KK mode number ($n_p = 100$). In Fig.~\ref{fig:singleFieldj6c5}, we show the effects of varying $\nu$. 
%these two parameters on the final-state event shapes predicted by the model. 
%We have studied samples with fixed jet multiplicity, to emphasize that the additional information captured by $\iso{sph}{192}$ is not simply a function of $\lmax$ and jet multiplicity. 
In the %left-hand panel, 
figure, we have plotted three values of $\nu$. As explained in~\cite{paper1}, larger values of $\nu$ lead to increased violation of KK number and less spherical events. This is captured by the shift of the contours toward larger values of $\iso{sph}{192}$. Notice that the distributions of $\lmax$ for the two cases $\nu = 0.15$ and $\nu = 0.75$ strongly overlap: $\iso{sph}{192}$ provides a means to resolve two models that are difficult to distinguish with thrust or the $C$ parameter. 
%In the right-hand panel, we show the evolution of the probability distributions with increasing $n_p$ in the case $\nu = 0$. Beginning higher up the decay chain leads to a larger final-state multiplicity and provides more opportunity for the event to become fully isotropic. We see that larger values of $n_p$ approach the theoretical lower bound on $\iso{sph}{192} \approx 0.1$ more closely, while $\lmax$ retains a wider range. A more detailed discussion of the dependence of event isotropy on $n_p$ may be found in Appendix B of \cite{paper1}.

%%%%%%%%%%%%%%%%%%%%%%%%%%%%%%%%%%%%%%%%%%%%%%%%
\subsection{Two field ($\nu_1 \neq \nu_2 = \nu_3$)}

We now consider the scenario of two scalar fields with coupling $\Phi_1 \Phi_2^2$ in the bulk. 
Whereas in the single field scenario the hadronic coupling constants decrease as KK-number violation increases, leading to KK-number near-conserving cascades, this is not true in the two-field case when  $\Delta \nu \equiv \nu_1 - 2 \nu_2 \geq 2$.  Instead there are ``plateaus'' in the space of  hadronic decay channels  where KK-number violation is large and the coupling constants are unsuppressed.
A cascade that proceeds through a plateau early can have large KK-number violation in early stages, producing boosted particles and leading to a much less isotropic radiation pattern than those of the single field scenario. 
As we increase $\nu_1$, the number of plateaus increases.
The amount of KK-number violation is strongly dependent on $\Delta \nu$ \cite{paper1}, so we consider the values $\nu_1 = 2, 3, 4$ but fix $\nu_2 = 0$. 
%
%
%%%%%%%%%%%%%%%%%%%%%%%%%%%%%%%%%%%%%%%%%%%%%%%%
\begin{figure}[!h]
\centering
\includegraphics[width=0.45\textwidth]{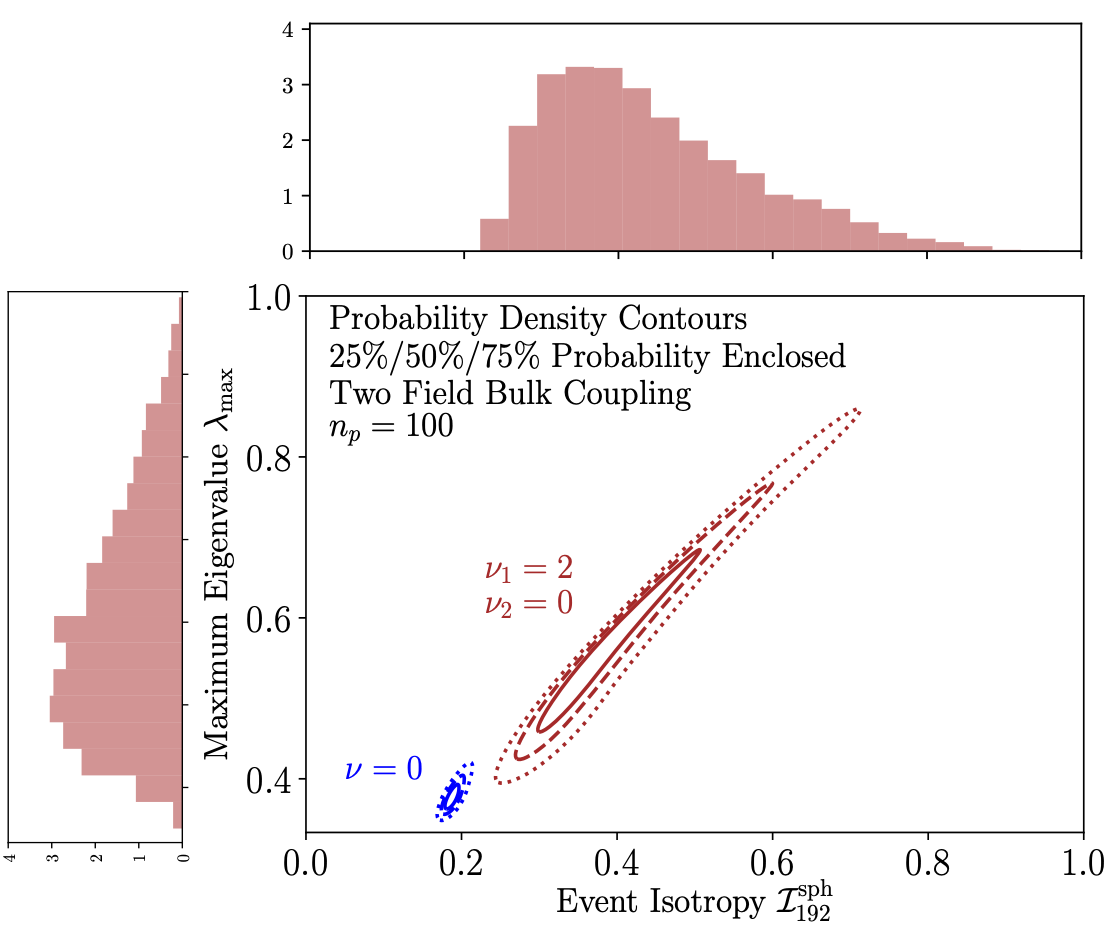}\ \qquad
\includegraphics[width=0.45\textwidth]{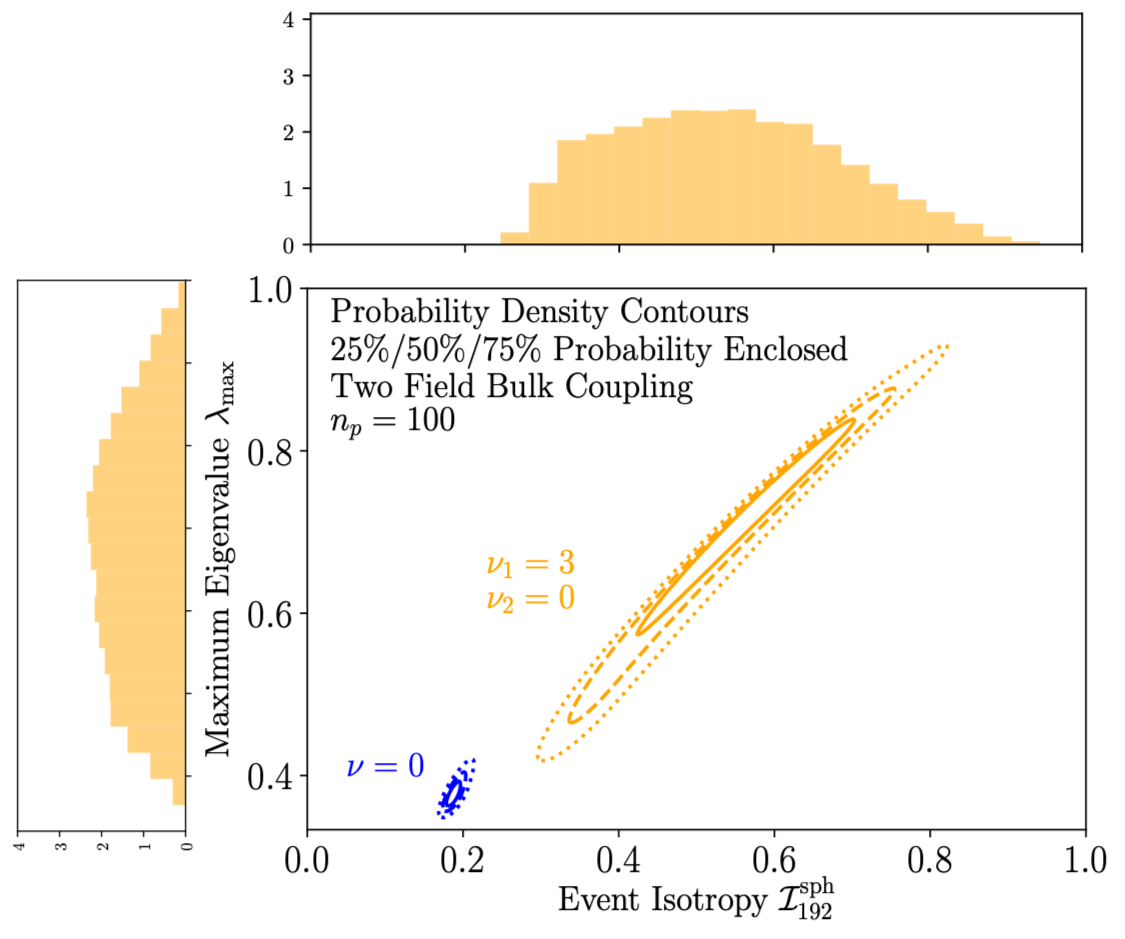}\
\hfill \newline \hfill \newline
\includegraphics[width=0.45\textwidth]{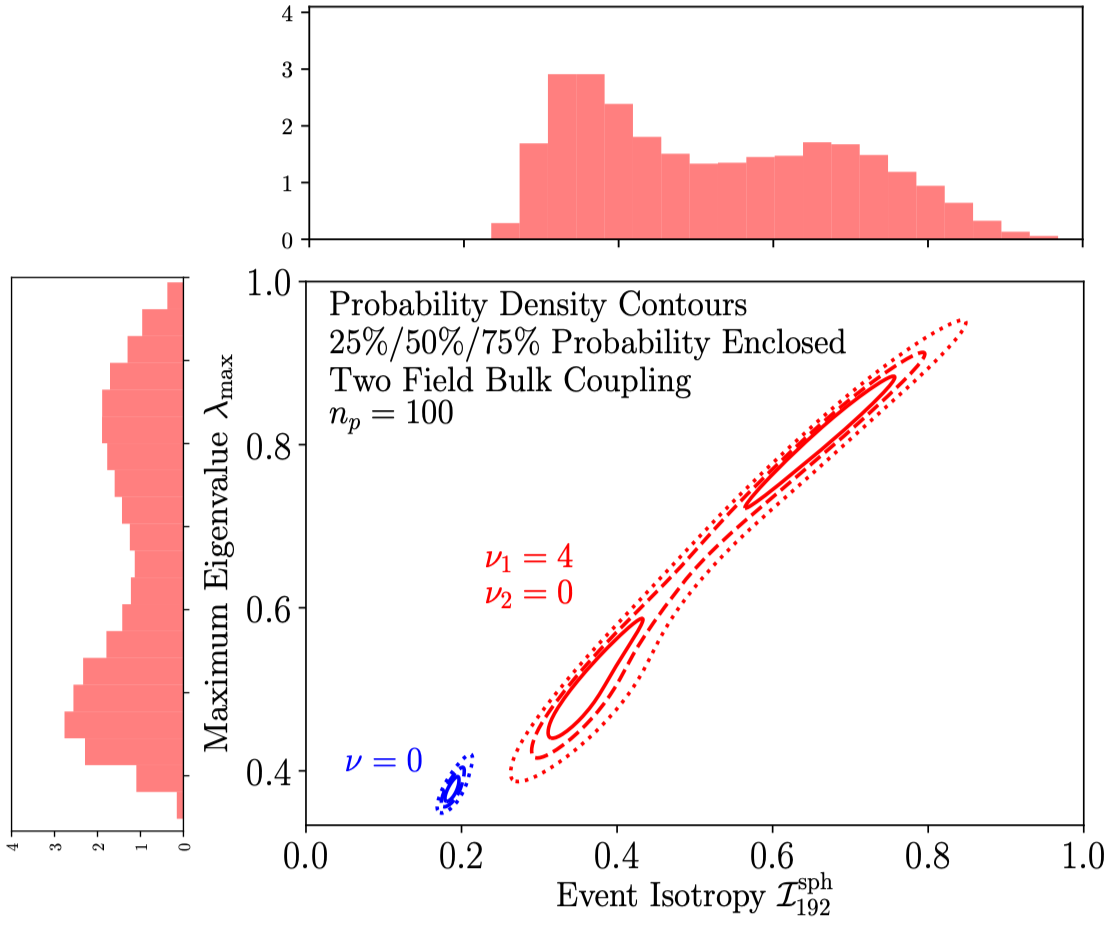}\
\caption{Contours of probability density quartiles (25\%, 50\%, and 75\% containment as solid, dashed, and dotted lines) as well as the projected distributions of $\iso{sph}{192}$ and $\lmax$ for our two-field simulations, which have $\nu_1=2,3,4$ with fixed $\nu_2=0$.  The $\nu=0$ single-field case is also shown for comparison.
}
\label{fig:twoFj5c5nu2}
\end{figure}
%
%%%%%%%%%%%%%%%%%%%%%%%%%%%%%%%%%%%%%%%%%%%%%%%%

We illustrate these effects in Fig.~\ref{fig:twoFj5c5nu2}, which shows the quartiles of probability density in the $(\iso{sph}{192}, \lmax)$ plane for the different choices of $\nu_1$. To allow easy comparison to the single-field case, each plot also shows the quartiles for the $\nu = 0$ single-field case. 
The most striking property we observe is the emergence of a bimodal structure in the $\nu_1 = 4$ example. 
%In the case of the $\iso{sph}{192}$ distribution, 
This was discussed in \cite{paper1}, and attributed to the distinction between events with decays on the plateau (with large boost) occurring early in the cascade and events where the initial decays are approximately KK-conserving. %The result in Fig.~\ref{fig:twoFj5c5nu2} reveals that this structure is visible even when we control for the number of jets in the event.

It is also interesting to note that the distributions for the two-field case lie at larger $\lmax$ than the distributions that appear in the $\nu=0$ single-field cascade evolution, Fig.~\ref{fig:fullNu0Cascade}; that is, for the same event isotropy, they have higher $\lmax$.  The particles in the single-field cases are more randomly oriented, leading to a sphericity tensor not far from unity and thus to smaller $\lmax$; meanwhile the boosted particles produced in a KK-number violating two-field cascade tend to be more correlated in angle, leading to larger $\lambda_{\rm max}$ for the same $N$ and $x$.  
 From the point of view of the estimator curves in \Fig{fig:earlyCasc} and \ref{fig:lateCasc}, this raises the distributions vertically and (because of the shape of the estimator curves) tilts them more horizontally. 
 
\subsection{Boundary couplings}

 With Neumann boundary conditions on the 5d wavefunctions, all of the KK mode wavefunctions have approximately equal values at $z_{\text{IR}}$.  As a result, a cascade through a $\Phi^3$ coupling localized entirely on the brane leads to branching ratios which are approximately given by the available phase space \cite{paper1}. Average violation of KK number is then large, so cascades are much shorter than for bulk $\Phi^3$ couplings, and the events are much less isotropic.  The number of jets that emerges becomes independent of the initial KK number $n_p$, because the boosted particles early in the cascade tend to determine the number of jets long before the cascade reaches its end. In particular, simulations of a field with $\nu = 0.3$ decaying through a boundary-localized $\Phi^3$ coupling give distributions in the ($\iso{sph}{192}, \lmax$) plane that look essentially identical for $n_p = 100, 300,$ and $500$.  This is in contrast to \Fig{fig:singleFieldNoJC}.
\begin{figure}[!h]
\centering
\includegraphics[width=0.5 \textwidth]{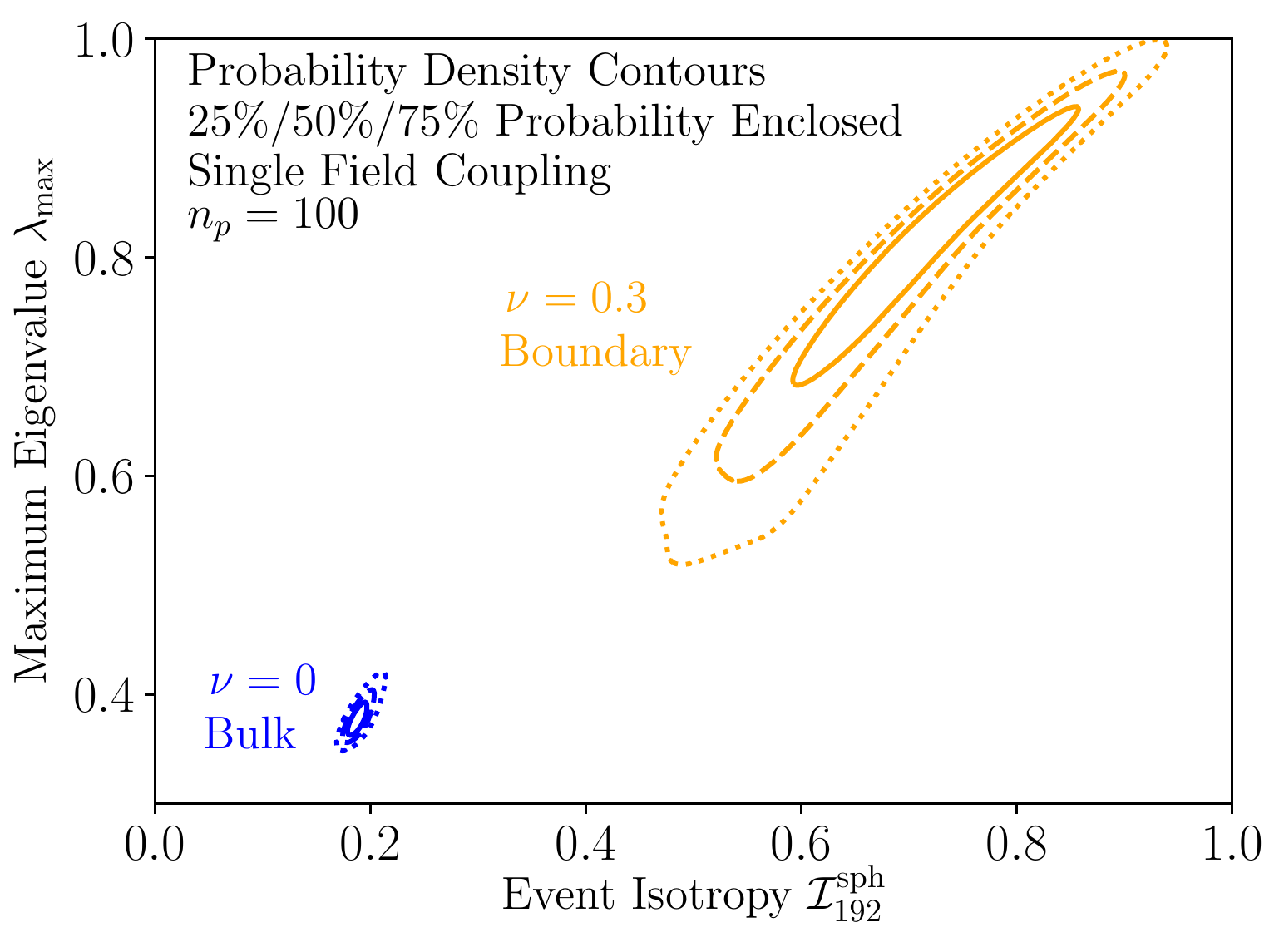}
\caption{Contours of probability density quartiles (25\%, 50\%, and 75\% containment as solid, dashed, and dotted lines) for final states in the single-field scenario, with bulk mass parameter $\nu=0.3$ and a boundary coupling; also shown is the benchmark case of $\nu=0$ with a bulk coupling. 
}
\label{fig:evtshapebulkvbdry2}
\end{figure}

In \Fig{fig:evtshapebulkvbdry2}, we focus on $n_p=100$ and plot probability density quartiles in the $(\mathcal{I}^\mathrm{sph}_{192}, \lambda_\mathrm{max})$  plane. 
%We control for jet multiplicity by plotting  only events with exactly six jets carrying more  than 5\% of the total energy in the event, a selection for which there are  still a substantial number of events in both samples  (see the lower panels of Fig.~\ref{fig:countsbulkvbdry}). The jet multiplicity selection removes the most dijet-like of the boundary events, preventing  the contours  from extending to the upper-right corner.
It is interesting to compare this with \Fig{fig:twoFj5c5nu2}; though the three two-field cases differ, they occupy narrow overlapping regions in the plane.  The boundary case is broader and extends to their right, deviating further from them at smaller event isotropy.  It turns out that because the near-threshold region is strongly disfavored by phase space for the boundary case, its events have much less energy in soft jets on average than do any of the two-field cases.  For example, for events with six hard jets, its soft energy is roughly what is found in step 3 in the $\nu=0$ cascade evolution, causing the contours in \Fig{fig:evtshapebulkvbdry2} to overlap those of step 3 in \Fig{fig:fullNu0Cascade}.

\begin{figure}[!h]
\centering
\includegraphics[width=0.5 \textwidth]{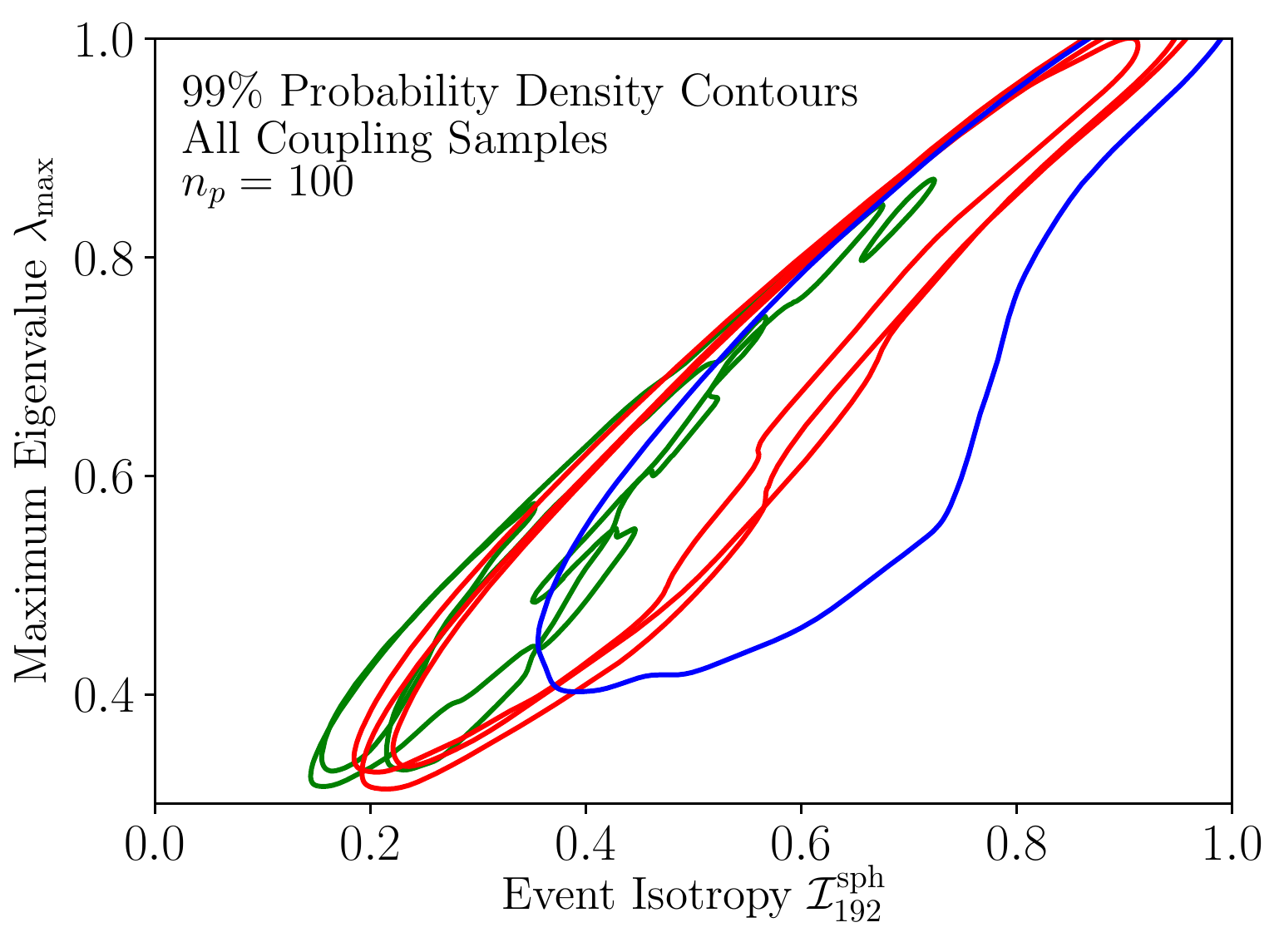}
\caption{
Contours in event isotropy $\iso{sph}{192}$ and maximum eigenvalue $\lambda_\text{max}$ that enclose 99\% of the events for all of our samples; single field samples are in green, two-field samples in red, and the boundary sample in blue.  Comparing with \Fig{fig:tvslambdamax}, we see that the samples are considerably less correlated than are $\lmax$ and thrust.
}
\label{fig:plotSummary}
\end{figure}

The difference between the two-field cases and the boundary case can be seen also in 
\Fig{fig:plotSummary}, which shows all of our simulations and uses $99\%$ contour lines, similar to those used in Figs.~\ref{fig:tvslambdamax}, \ref{fig:DvsLmax}, and \ref{fig:mults}.  This shows that the single-field cases occupy one region of the plane, the two-field cases share a different but overlapping region, and the boundary case occupies a third that overlaps the two-field region but not the single-field region.  While it is no surprise that the quasi-spherical single-field cases would be distinct from the others, the fact that event isotropy and $\lmax$ {\it combined} (but not separately) can distinguish these various classes of complex signals is significant.

%%%%%%%%%%%%%%%%%%%%%%%%%%%%%%%%%%%%%%%%%%%%%%%%%%%%%%%%%
\section{Discussion}
\label{sec:discussion}

In order to sift through the remarkable amount of data collected at the LHC and other high luminosity colliders for new physics signals, it is essential to have as many uncorrelated handles as possible when analyzing data. 
Since many event shape observables have been designed to probe deviations from an underlying two-parton QCD event, there is motivation to develop observables that are sensitive to signals that differ greatly from QCD-like events, such as (but not limited to) spherical radiation patterns. 
While certain older observables such as thrust and eigenvalues of the sphericity tensor are highly correlated, as shown in \Fig{fig:tvslambdamax}, and are most sensitive to QCD-like events, event isotropy provides useful new information and is more effective at separating quasi-spherical events from jetty events.

This is especially clear when we examine distributions in event isotropy and $\lmax$, the largest eigenvalue of the sphericity tensor.
% (which is highly correlated with thrust and with the $C$ parameter.)
%
In \Fig{fig:fullNu0Cascade} (whose contours show probability quartiles, not standard deviations),  it can be seen that the evolution of a cascade from jetty to spherical spreads events across a large portion of the plane. 
Meanwhile \Fig{fig:plotSummary}, whose contours contain $99\%$ of the events from multiple simulations with very different phenomenology, show that physically reasonable events can populate large portions of the plane in a way that is usefully model-dependent.
Moreover, we have seen in \Fig{fig:mults} that event isotropy does not merely count particles or jets, and therefore it gives information complementary to these counting measures.

Much more remains to do.
 We have not performed a comprehensive study of existing event-shape observables, and there are others  that should be compared with event isotropy to see which ones are complementary and which ones redundant.
In this regard it may be useful to view such observables in terms of their EMD interpretation, where known \cite{Komiske:2020qhg}.

In addition, our study is not yet directly applicable to experimental investigations.  
Little is currently known about either complex hidden-valley-type signals or about event isotropy, so for clarity of analysis we have limited ourselves to background-free signals in the context of a spherical geometry, far from the world of hadron colliders.
Moreover, again for clarity and simplicity, we have analyzed a toy model of a hidden valley with an incompletely defined signal; for instance, the coupling of the hidden sector to the Standard Model was not specified, so the production process and detector-level final state are not determined.
Nevertheless, one advantage of event-shape observables is that they are sensitive to the flow of energy, and not to the particles that carry them, as long as those particles are detectable, and thus they are less model-dependent than many other observables.
We hope therefore that the lessons we have learned are general, and will be applicable and useful in more realistic studies of complex signals.
%
%Since it is possible that our community is  overlooking such signals in existing data, more research on this topic is important.

\section*{Acknowledgments}

We thank Marat Freytsis, Patrick Komiske, Eric Metodiev, Gavin Salam, and Jesse Thaler for useful conversations. MR and CC are supported in part by the DOE Grant DE-SC0013607. CC is supported in part by an NSF Graduate Research Fellowship Grant DGE1745303. We have made use of the FastJet \cite{Cacciari:2011ma}, GetDist \cite{Lewis:2019xzd}, SciPy \cite{2020SciPy-NMeth}, and Python Optimal Transport \cite{flamary2017pot} packages. 

\appendix

\section{Finite-Multiplicity Effects on Event Isotropy Calculations}
\label{ap:EMD}

\subsection{Event Isotropy of Quasi-Uniform Radiation Pattern}

Samples of events with finite multiplicity $k$, when compared to a reference sphere of $\kref$ isotropically distributed pixels, have an intrinsic floor to their average event isotropy. 
We derive this theoretical minimum here. 
Details of this calculation were originally presented in \cite{Cesarotti:2020hwb}.

Discretized isotropic distributions $\mathcal{U}^\text{sph}_k$ of $k$ particles on a sphere can be generated using  \texttt{HEALPix} \cite{2005ApJ...622..759G}.
Only specific values of $k$ are allowed:
\be
k  = 12 \times 2^{2i}, \qquad \qquad i \geq 0.
\ee
What we want to do is calculate the event isotropy of a spherical event with $k$ massless symmetrically distributed particles to a $\kref$-pixel reference sphere. 
Clearly, if $k=\kref$, the event and the reference sphere could be chosen to be perfectly aligned, so the minimum event isotropy per event is zero.
But what is of more interest is the {\it average} event isotropy obtained if we rotate the spherical event randomly relative to the reference sphere.
This quantity is only zero for $k,\kref\to\infty$.

%
%
%The event isotropy is only identically zero for all orientations when we take the number of particles in both distributions to be infinite. 
%%
%The averaged event isotropy of any finite particle radiation pattern therefore has a minimum non-zero value. 
%%
%As a baseline we consider finite multiplicity events consisting of massless particles with equal momentum and a uniform distribution in solid angle.
%%
%To quantify how spherical the events from our simplified 5d model are, we compare to this baseline instead of the extremal value $\iso{}{} =0$.
%
%%
%In this section we provide an analytic estimate for the minimal value of $\iso{sph}{}$ for a finite particle distribution.
%%
%We assume $n \rightarrow \infty$ whereas $k$ remains finite.
%

Let us consider the case $\hat k=\infty$, $k$ finite; that is, the event is in the form of equal-energy particles distributed as a \texttt{HEALPix} distribution with $k$ points, and the reference sphere is smooth.

In this case the energy of each particle in the event must be uniformly distributed over a region centered at the particle, which has area
\be
A_k = \frac{4\pi}{k}.
\label{eq:areaTiling}
\ee
%

%%%%%%%%%%%%%%%%%%%%%%%%%%%%%%%%%%%%%%%%%%%%%%%%
\begin{figure}[!h]
\centering
\includegraphics[trim = 0 10 30 35, clip, width=0.45\textwidth]{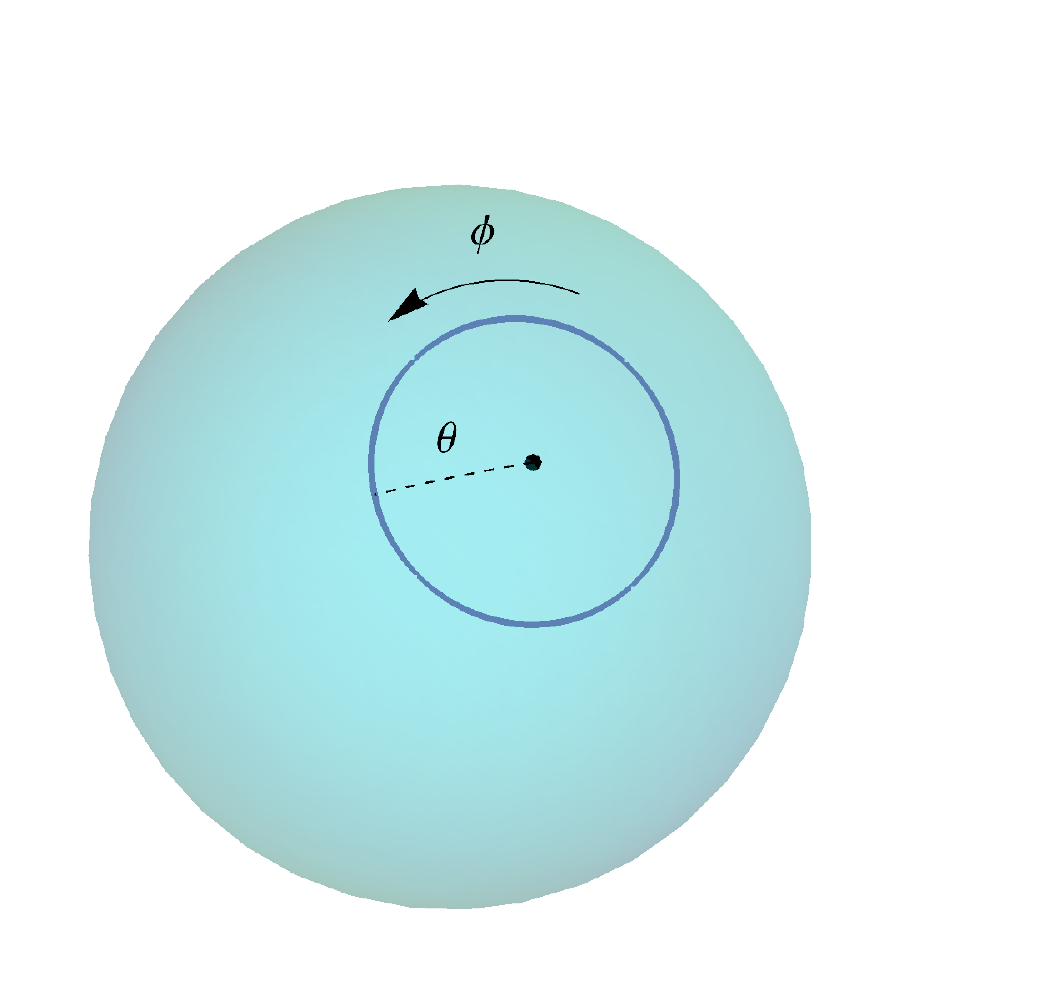}\
\caption{An illustration of a single particle on the sphere. We approximate the region over which the energy is spread as a disk. The radius of the disk is labeled by the polar angle from the center $\theta$, and the circumference of the disk is parameterized by $\phi$.}
\label{fig:tiling}
\end{figure}
%
%%%%%%%%%%%%%%%%%%%%%%%%%%%%%%%%%%%%%%%%%%%%%%%%
%

\begin{figure}[t!]
\centering
\includegraphics[width=0.55\textwidth]{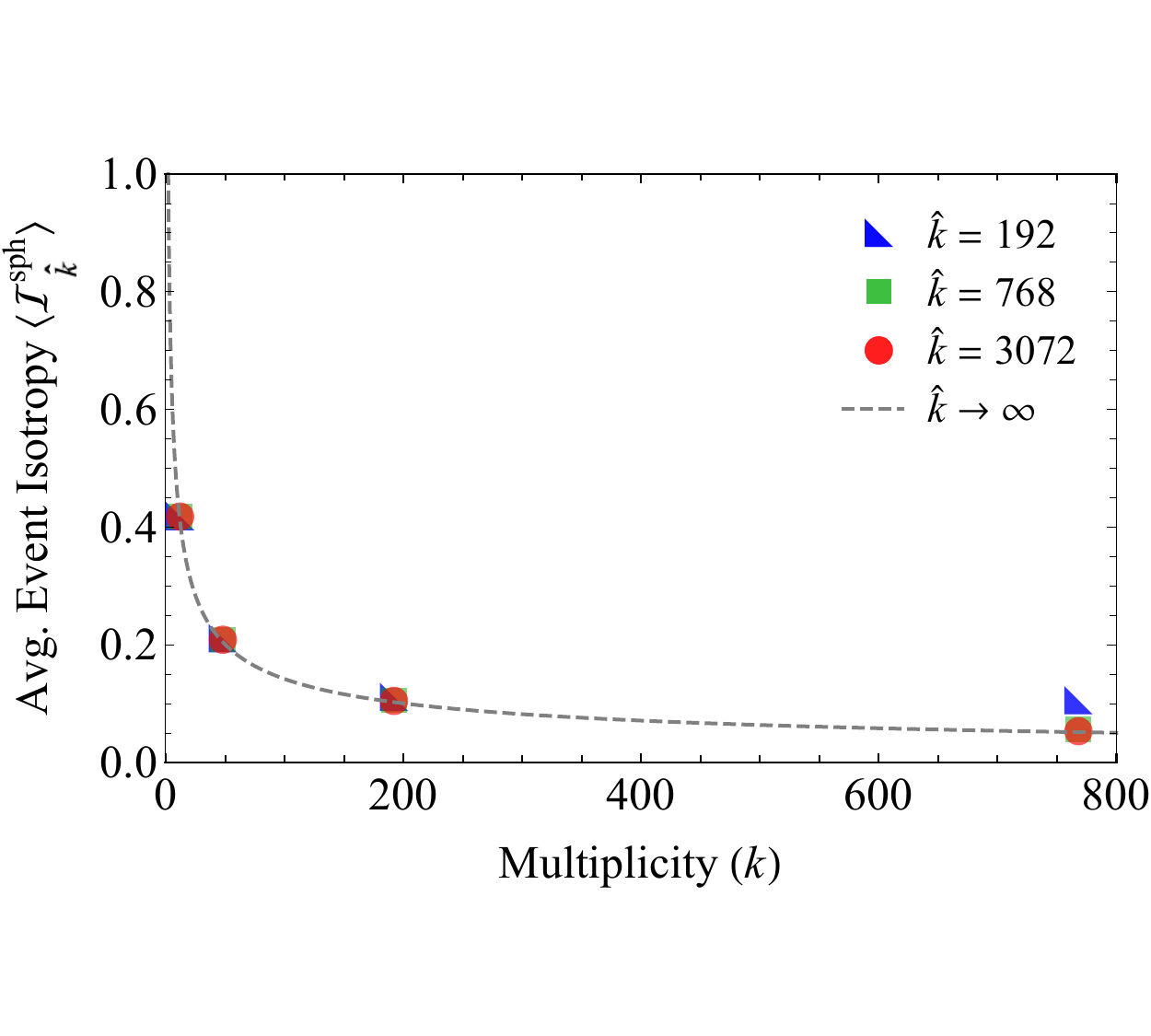}
\caption{ A comparison of the event isotropy $\iso{sph}{\kref}$ for $\kref=192, \ 768$, and $3072$ to quasi-isotropic distributions of $k$ particles. The solid markers indicate explicit values calculated from averaging over orientations of \texttt{HEALPix} events. The dashed line is the approximation from \Eq{eq:approx0}.}
\label{fig:healpixAvg}
\end{figure}

For $k$ sufficiently large we can approximate these patches as disks on the surface of a sphere (see \Fig{fig:tiling}).
From \Eq{eq:areaTiling}, the radius of these patches is the maximum polar angle drawn from the center of the disc, approximately 
\be
\theta_k \approx \frac{2}{\sqrt{k}}
\ee
Using the distance measure in \Eq{eq:distMeas}, one finds 
\be \begin{aligned}
\iso{sph}{\infty}\left( \mathcal{U}_k\right) & \approx \frac{1}{A_k} \times \int_0^{\theta_k} \int_0^{2\pi} \theta \,{\rm d}\phi \,{\rm d}\theta\,\frac{3}{2} \sqrt{1-\cos \theta} \\
& \approx \frac{1}{A_k} \times \int_0^{\theta_k} \int_0^{2\pi} \theta \, {\rm d}\phi \,{\rm d}\theta \ \frac{3}{2} \sqrt{\frac{\theta^2}{2}} \qquad  ( \theta \ll 1) \\
& = \sqrt{\frac{2}{k}}  \ 
\end{aligned} 
\label{eq:approx0}
\ee
(Note that this result is correctly normalized: for $k=2$, with two back-to-back particles, the event isotropy should be 1.)
In \Fig{fig:healpixAvg} we compare this approximation to numerical calculation for finite  $k$ and $\kref$; we see the theoretical curve matches the case when $k\ll \kref$. 

In the opposite limit, where $k\gg \kref$, the result becomes independent of $k$.  This is easy to understand; the EMD is symmetric under exchange of the event and the reference sphere, so the calculation of the average EMD for two \texttt{HEALPix} spheres is symmetric under $k\leftrightarrow\kref$. Thus, for finite $\kref$, we have
\be
\iso{sph}{\kref}\left( \mathcal{U}_k\right) \approx \sqrt{\frac{2}{\text{min}(k,\kref)}}  
\label{eq:approx}
\ee

Since the multiplicity of the cascades in this paper is typically on the order of $\mathcal{O}(100)$, we do not expect to find $\iso{sph}{\kref}<\sqrt{2/200}$ in any of our results, no matter what we choose for $\kref$.
We therefore take $\kref=192$, and measure the degree to which an event sample is spherical by comparing its event isotropy to the theoretical minimum  $\sqrt{2/192}\sim 0.1$.

\subsection{Event Isotropy of a Multiprong Plus Spherical Event}

In \Sec{sec:multiprongplussphere}, we discussed a class of events that combine a multiprong event carrying a fraction $x$ of the energy with a soft spherical event carrying a fraction $(1-x)$ of the energy. In this case, when both the soft sphere and the reference sphere are perfectly uniform $(k=\kref=\infty$), the event isotropy took the simple form of  \Eq{eq:addsphere}. However, discretization effects modify this calculation. If the finite spheres have equal particle number $k=\kref$ and are perfectly aligned, there is no need to reorganize the soft sphere, and so no EMD cost.   The EMD is then computed by spreading the energy in the multiprong component to the reference sphere, and \Eq{eq:addsphere} still holds. But if the soft and reference spheres are not aligned, energy in the spherical component must be moved.  Thus we expect \Eq{eq:addsphere} to provide a lower bound on the  event isotropy, as in \Eq{eq:addsphererange}.

When both the multiprong component and the spherical component of the event must be rearranged, the optimal transport map $f_{ij}$ appearing in \Eq{eq:EMDdef} is no longer determined by separately rearranging the two components. It becomes more efficient, especially at small $x$, to reorganize the multiprong event's energy more locally, rather than spreading it evenly over an entire half-wedge as in \Fig{fig:slice}. As a result, the true event isotropy will be smaller than the upper bound determined by the convexity property \eqref{eq:EMDconvexity}.

These expectations lead us to a finite range in which we expect the event isotropy to lie. When the soft sphere in the event has $k$ particles and the reference sphere has $\kref$ pixels, and they are misaligned by a generic angle, we expect the EMD contribution from the finite spheres to be given by \Eq{eq:approx}.
The event isotropy of the combined event with the multiprong and spherical components is then expected to lie in the interval 
\begin{equation}
x \iso{sph}{\kref}(x=1) \leq \iso{sph}{\kref}(x) \lesssim x\iso{sph}{\kref}(x=1)+ (1-x)\sqrt{2/{\rm min}(k,\kref)},
  \label{eq:isosphapprox}
\end{equation}
when averaging over the relative orientation of the soft sphere and the reference sphere.

\begin{figure}
\centering
\includegraphics[width=0.65\textwidth]{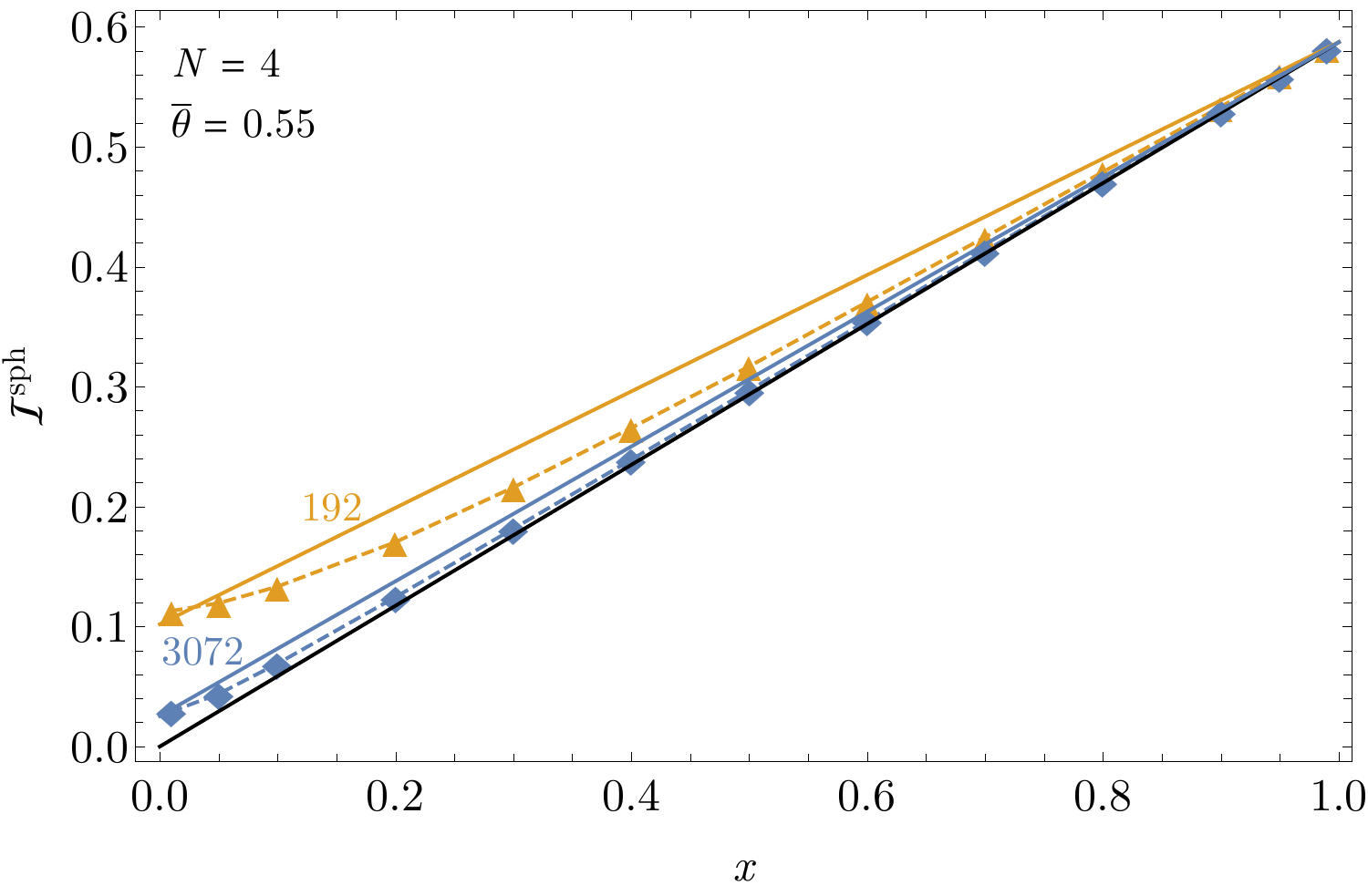}
\caption{ Event isotropy $\iso{sph}{\kref}$ as a function of multiprong energy fraction $x$. We have fixed the parameters of the multiprong event to be $N = 4$ and $\bar \theta = 0.55$. The solid lines correspond to the lower bound (black) and upper bounds (orange for $k = \kref = 192$, blue for $k = \kref = 3072$) from \Eq{eq:isosphapprox}. The dashed curves connect computed data points, %The lower dashed curves, which lie on top of the black line, are for the case where the spherical component of the event is perfectly aligned with the reference sphere (but we average over a random orientation of the multiprong event relative to this sphere).  The upper dashed curves, which 
and lie  between the upper and lower bounds for the corresponding choice of $k$.}
\label{fig:eviso_xdep}
\end{figure}

 To explore this effect, in \Fig{fig:eviso_xdep} we compare the average computed event isotropy to the lower and upper bounds in \Eq{eq:isosphapprox}. For two different choices of $k$ (taking $\kref = k$ for simplicity), we take samples of events in which the soft sphere is randomly oriented relative to the multiprong particles, and these are in turn rotated randomly relative to the reference sphere.  The average  event  isotropy lies strictly between the expected lower and upper bounds, with the spherical contribution becoming of increasing importance as $x$ shrinks toward 0. As $k=\kref$ becomes arbitrarily large, the upper bound collapses onto  the lower bound, confirming that \Eq{eq:addsphere} is exact in the limit of $k,\kref\to\infty$. 

%%%%%%%%%%%%%%%%%%%%%%%%%%%%%%%%%%%%%%%%%%%%%%

\section{Adding a Dijet to the Estimator}
\label{app:AddDijet}

We can expand the domain of our estimator to include the following events.  To our $2N$ vectors $\vec p_n$, $n=1\dots 2N$ defined in \Eq{eq:vecDef}, we add vectors $q\hat z$ and $-q\hat z$ ($q>0$).  The sphericity tensor remains diagonal but the maximum eigenvalue is sensitive to the additional dijet. From \Eq{eq:spherTensor},
\be
\lmax = \max\left(\frac{q + p\cos^2\bar\theta}{q+p}, \frac{p  \sin^2 \bar\theta}{2(q+p)}\right).
\label{eq:lambdaMPD}
\ee
The critical angle beyond which the $\lmax$ eigenvector no longer points in the $z$ direction is determined by
\be
\cos \theta_\text{crit} = \sqrt{\frac{1}{3}  - \frac{2q}{3p}}.
   \label{eq:thetacritgeneral}
\ee
%When  $q =  0$, we recover our earlier restriction $\cos\bar \theta > 1/\sqrt{3}$. 

By adjusting the ratio $q/p$ and $\bar\theta$, we can move around on the set of configurations with a fixed $\lambda_\text{max}$. For $\bar \theta <  \theta_\text{crit}$, inverting \Eq{eq:lambdaMPD}, we have
\be
0 < \frac{q}{p} = \frac{\lmax-\cos^2\bar\theta}{1-\lmax}  < \frac{\lmax}{1-\lmax} \ .
\ee
Note as we turn off the dijet, i.e., $q/p\to 0$, the maximum eigenvalue $\lambda_\text{max}$ simplifies to $\lmax=\cos^2\bar\theta$, as above. If we take the pure dijet limit $q/p\to\infty$, sensibly we have  $\lmax=1$.

The geometry of this configuration makes the event isotropy more difficult to estimate. At large $N$ and small $q/p$, the vector pointing along the $z$ axis will have its energy spread onto a cap, approximately a disk surrounding the north pole of the angular sphere, while the remaining vectors in the positive $z$ hemisphere will again spread their energy in equal azimuthal slices of the hemisphere with the above-mentioned cap removed.  This spreading should lead to a constant energy density, so the cap, extending from $\theta=0$ down to $\theta_c$, should be chosen so that its area, relative to the area of each truncated half wedge, is $q/(p/N)$.  This requires
\be
2\pi  (1-\cos\theta_c) /q= \frac{2 \pi}{N} \cos\theta_c/(p/N)
\label{eq:newAngleBounds}
\ee
and thus $\cos\theta_c=p/(p+q)$.  
Our approximation can only be plausible  when $\cos\theta_c>\cos\bar\theta$, so that the cap does not enclose the vectors $\vec p_j$. 

Under these assumptions,
\bea\label{eq:withpolarcap}
{\cal I} &\approx& \frac{1}{4\pi/3} (2 \pi)  \int_{p/(p+q)}^1 {\rm d}\cos\theta \ \sqrt{1 - \cos\theta}
\nonumber\\
& & + \frac{N}{4\pi/3} \int_0^{p/(p+q)} {\rm d}\cos\theta\ \int_{-\frac{\pi}{N}}^{\frac{\pi}{N}}\  {\rm d}\phi \
\sqrt{1 - \hat n_0\cdot \hat r}
\nonumber\\
&=&\left(\frac{q}{p+q}\right)^{3/2} + 
\frac{3N}{\pi} \int_0^{p/(p+q)} {\rm d}\cos\theta \ \sqrt{1 - \hat n_0\cdot \hat r_0} \ E\left(\frac{\pi}{2N},-\frac{2 \sin \theta \sin \bar \theta}{1 - \hat n_0\cdot \hat r_0}\right)
 \ ,
\eea
where $\hat n_0, \hat r, \hat r_0$ are defined as in \Eq{eq:Isotropy2N}.

\begin{figure}[!h]
\centering
\includegraphics[width=0.45\textwidth]{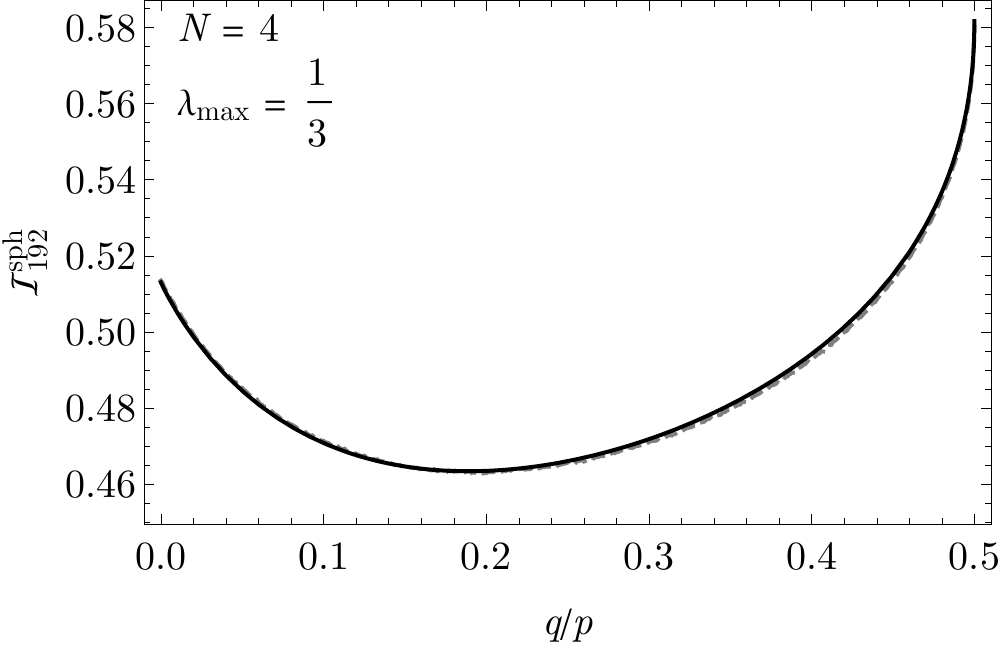}
\caption{Event isotropy $\iso{sph}{192}$ for events with $N = 4$ and $\lmax =  1/3$, varying the dijet ratio $q/p$. The solid curve is \eqref{eq:withpolarcap} and the dashed curve is a numerical calculation, averaging over orientations of the 192-pixel \texttt{HEALPix} reference sphere.}
\label{fig:isoqoverp}
\end{figure}

Again we may make some simple estimates before analyzing these equations, starting with $N=4$ and $\lmax=1/3$.  When $q=0$, we have eight isotropically distributed vectors and ${\cal I}\sim 1/\sqrt{4}$.  When $q/p=1/4$, we have ten vectors of the same energy, roughly isotropically distributed, so we expect that ${\cal I}\sim 1/\sqrt{5}\sim 0.45$.  When $q/p=1/2$, its maximum value where $\cos\bar\theta=0$, the eight vectors $p_j$  reach the equator and merge into four, giving a total of six equal vectors along the positive and negative $x,y,z$ axes.  For this arrangement we expect ${\cal I} \sim 1/\sqrt{3} \sim 0.58$.  Thus we expect the isotropy is a non-monotonic function of $q/p$, first decreasing and then increasing by of order 10\%.    This behavior, shown in Fig.~\ref{fig:isoqoverp}, is also clear from the equations:  for small $q/p$, the first term in (\ref{eq:withpolarcap}) initially grows as $(q/p)^{3/2}$ while the second decreases linearly in $q/p$.  This decrease continues briefly until a minimum is reached around $q/p \sim 0.2$, almost independent of $N$. 

The approximation that the $q\hat z$ vectors spread their energy into a simple polar cap  must fail as $\cos\bar\theta\to 0$ and $q/p \to \lmax/(1-\lmax)$, because a better approximation is that the two polar vectors spread their energy over the entire hemisphere except for little caps surrounding the other (now equatorial) vectors.  However, numerically it turns out that the region in which this is important is quite small, and the numerical calculation of the isotropy agrees quite well with (\ref{eq:withpolarcap}).   Furthermore, this subtlety is largely irrelevant, as phase space disfavors events that probe the region of maximum $q/p$ for fixed $\lmax$.

\begin{figure}
\centering
\includegraphics[width=0.65\textwidth]{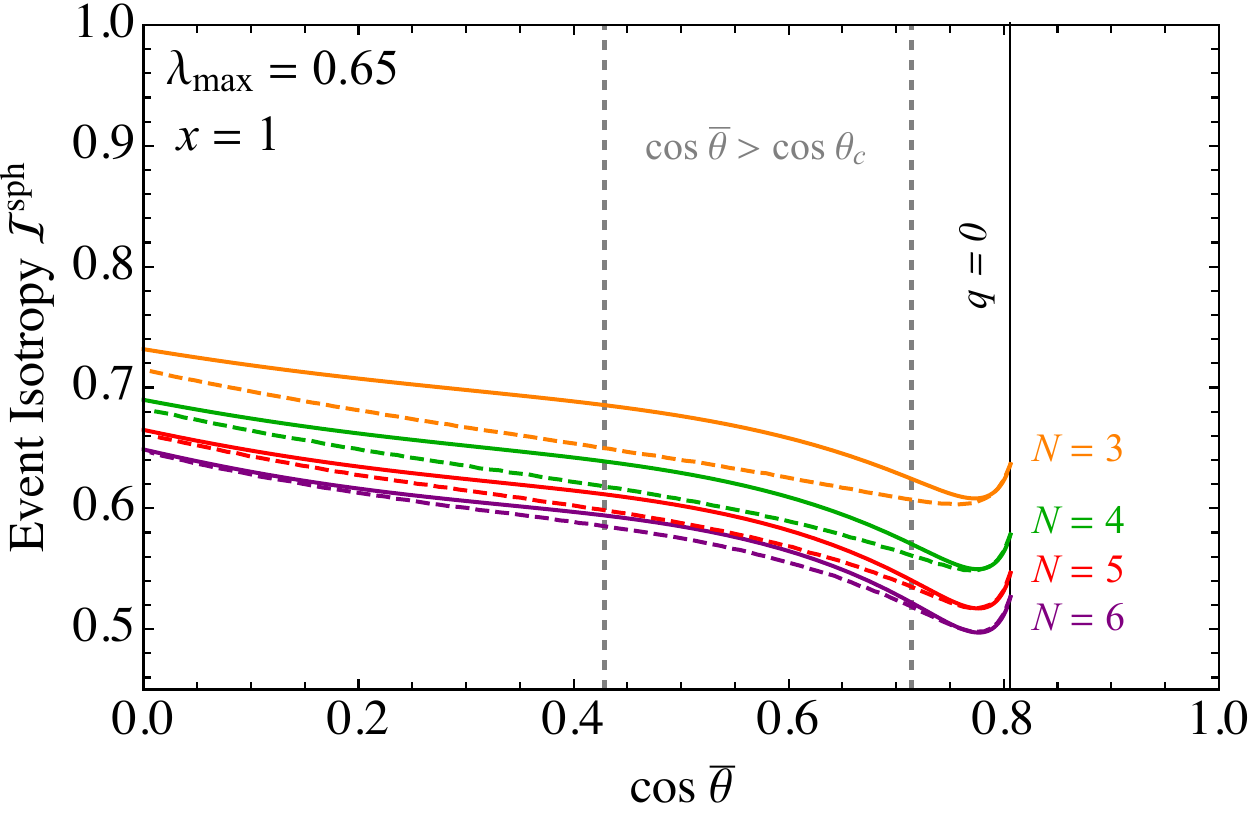}
\caption{ For fixed values of $\lambda_\text{max}$ and $x$, the event isotropy  for different choices of $N$ and $\cos \bar{\theta}$. We fix $\lambda_\text{max} = 0.65$ (\Eq{eq:lambdaMPD}) and scan over the valid range of $\cos\bar{\theta}$, adjusting $q/p$ to keep $\lambda_\text{max}$ constant. The different curves correspond to different choices of $N$.  Solid curves are the analytic formula \eqref{eq:withpolarcap}, whereas dashed curves are numerical calculations averaging over orientations of the 192-pixel \texttt{HEALPix} sphere. The plot cuts off at $\cos {\bar \theta} = \sqrt{\lmax}$, where $q = 0$. We highlight the region bounded by the gray dashed lines, where $\cos\bar\theta > \cos\theta_c$, as it is outside the range of our approximation.}
\label{fig:evIsoVary}
\end{figure}

The conclusion is that, for fixed $N\sim 3-5$ and fixed $\lmax\sim 1/3-1/2$, varying $q/p$ can cause event isotropy to range of order 0.05 above or below the value at $q/p=0$.  This range decreases roughly linearly with $\lmax$ as $\lmax \to 1$, though in practice phase space and the constraint in \eqref{eq:newAngleBounds} restrict it further.

We now examine the approximation of \Eq{eq:withpolarcap} at larger $\lmax$. In this regime, a majority of the momentum is oriented along the $z$-axis and the polar cap approximation will likely break down. For concreteness we consider $\lmax = 0.65$, as shown in \Fig{fig:evIsoVary}. The deviation between computed event isotropy and theory curves from \Eq{eq:withpolarcap} is most significant in the regime where the polar cap extends beyond $\bar\theta$ (delineated by dashed vertical lines), where we did not expect the approximation to apply. The overall agreement between computation and theory improves for larger $N$, as the azimuthal symmetry is enhanced. 

What this figure shows is that the approximations above correctly capture basic facts: that even for a fixed choice of $\lmax$, event isotropy can vary by amounts of order 10\% as $\bar \theta$ is varied over its allowed range while holding $N$ fixed, and that the $N$ dependence is also a  10\%-20\% effect.% Thus, it is evident that the $\lmax$ and $\iso{sph}{}$ observables do not provide redundant information about the event shape.  \ms{Is it evident? and didn't we say the same thing two paragraphs ago?  Not sure how to conclude here, but this isn't the right way yet.}

\bibliography{ref}
\bibliographystyle{utphys}

\end{document}